\documentclass{svjour3}

\usepackage{enumitem}
\usepackage{hyperref}
\usepackage{tabularx}
\usepackage{xurl}  
\usepackage{graphicx}
\usepackage{amsmath}
\usepackage{hyperref}
\usepackage{multirow, multicol, makecell}
\usepackage[breakable,skins]{tcolorbox}
\usepackage{svg}
\usepackage{xcolor}
\usepackage{listings}

\newcommand{\todo}[1]{\textcolor{black}{ #1}}
\newcommand{\update}[1]{\textcolor{black}{ #1}}

\newcommand{\knowledge}[1]{{Knowledge \& Concept Support}}

\newenvironment{tightquote}
  {\list{}{\leftmargin=1.5em
            \rightmargin=1.5em
            \topsep=0pt
            \parsep=0pt
            \itemsep=0pt}%
   \item\relax\color{black!80}}
  {\endlist}

\begin{document}
\sloppy


\title{Empirical Study on the Characteristics and Evolution of AI-usage in GitHub Repositories: Evidence from Code Comments}

\author{Abdullah Al Mujahid \and Preetha Chatterjee \and Mia Mohammad Imran}

\institute{
  Abdullah Al Mujahid\at
  Missouri University of Science and Technology \\
  \email{amgzc@mst.edu}
  \and
  Preetha Chatterjee\at
  Drexel University \\
  \email{preetha.chatterjee@drexel.edu}
  \and
  Mia Mohammad Imran\at
  Missouri University of Science and Technology \\
  \email{imranm@mst.edu}
}

\maketitle

\begin{abstract}

\textbf{\\Context:} Developers are increasingly incorporating AI tools such as ChatGPT, Copilot, and Claude into everyday workflows. Prior studies often evaluate LLM outputs in isolation, leaving underexplored how developers adapt and integrate these suggestions into real-world projects.

\noindent \textbf{Objective:} We analyze 35,361 GitHub code comments that explicitly reference AI use and associated code blocks, examining supported tasks, post-introduction revisions, and temporal usage patterns.

\noindent \textbf{Method:} We open-code 500 unique code comments with their associated code blocks to derive a taxonomy of AI-assisted development activities. We then annotate 35,361 comments using two LLM-based classifiers, aggregating their predictions through the Dawid-Skene Expectation-Maximization framework (\textbf{RQ1}). To examine how AI-assisted code evolves after introduction, we analyze 12,996 commit messages associated with subsequent changes (\textbf{RQ2}). Finally, we conduct a longitudinal analysis of these 35,361 AI-referencing code comments and associated code blocks from December 2022 to March 2026 (\textbf{RQ3}).

\noindent \textbf{Results:} We find that developers primarily use LLMs for \textit{code implementation}, followed by \textit{code enhancement}, \textit{debugging}, \textit{documentation}, and \textit{testing}. Developers use LLMs as collaborative tools to generate new ideas, explore alternatives, and refine solutions. The 12,996 subsequent commit messages frequently involve \textit{refactoring \& cleanup}, \textit{feature integration \& extension}, and \textit{bug fixing \& corrective changes}, indicating sustained human oversight in adapting AI-assisted code to real-world projects. Longitudinally, AI-referencing comments and associated code blocks show a shift from direct code generation toward greater emphasis on knowledge and conceptual support, and code enhancement, reflecting the evolving integration of AI into everyday software development.

\noindent \textbf{Conclusions:} AI tools are becoming embedded in software development not only as a code-generation aid, but also as a collaborative support mechanism whose outputs are refined, extended, and corrected by developers over time.
\end{abstract}

\section{Introduction}

\update{AI tools and} large language models (LLMs) such as GitHub Copilot, ChatGPT, and Claude have become integral to modern software development. Developers now frequently rely on these models to generate code, explain~\cite{barke2023grounded}, refactor existing modules~\cite{alomar2024refactor}, and produce documentation
~\cite{dvivedi2024comparative}. This widespread adoption is reshaping the nature of programming, from an exclusively human activity to a collaborative process involving both human and AI contributions~\cite{murali2024ai}. As these tools become embedded within daily workflows, understanding how developers actually use \update{AI} in practice has become an essential question for software engineering research~\cite{mozannar2024reading}.

Most prior studies on \update{AI}-assisted programming rely on controlled experiments or benchmark evaluations to measure model performance in code generation and related tasks~\cite{peng2023impact,mozannar2024reading,barke2023grounded}. 
Benchmark-based and user studies have analyzed tools such as GitHub Copilot, Claude, and ChatGPT, as well as models like Llama, Qwen, and Gemini, primarily focusing on developer productivity, usability, code quality, error correction, and problem-solving processes~\cite{du2024evaluating,jimenez2023swe}. While these studies offer useful technical insights, they largely reflect artificial or short-term evaluation settings. How developers actually use, adapt, and integrate \update{AI} outputs in real projects remains largely unexplored, particularly how such code changes and matures once integrated into software repositories.


To address this gap, we collect and examine artifacts from real GitHub projects, specifically those that reveal how large language models are used within development workflows. 
By focusing on these in-situ artifacts, we can observe how developers employ, interpret, and modify \update{AI}-generated outputs in different programming or task contexts. 
Although prior work has examined commits, code reviews, pull requests, and issue discussions~\cite{chouchen2024software,grewal2024analyzing,guo2024exploring,ehsani2025towards,Ehsani_EMSE}, these efforts mainly capture higher-level collaboration and coordination activities. The finer-grained interactions, i.e., how developers engage with and adapt model-generated code within source files, remain comparatively underexplored. 

\begin{figure}
    \centering
    \includegraphics[width=\linewidth]{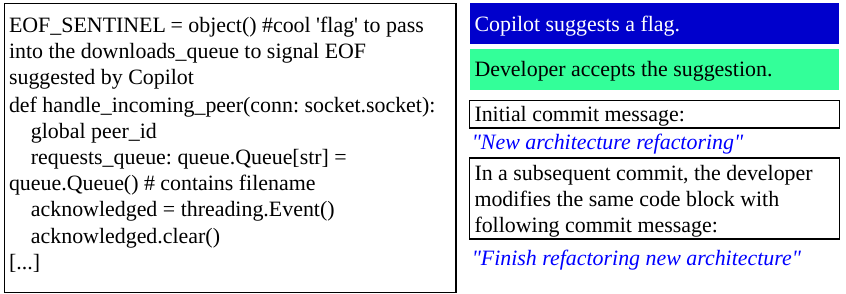}
    \caption{Example of a Real-World Code Comment and Connected Commits Reflecting Human-AI Collaboration.}
    \label{fig:comments-example}
\end{figure}

Among these finer-grained artifacts, code comments provide a particularly valuable lens for analysis~\cite{rani2023decade}. 
Comments capture developers’ reasoning in context: they describe intent, limitations, and uncertainty, and often include self-admitted acknowledgments of model involvement (e.g., ``\texttt{generated by ChatGPT},'' ``\texttt{suggested by Copilot},'' ``\texttt{AI-generated, please review}''). Unlike commits or documentation, comments are co-located with code, offering fine-grained evidence of how developers interpret and modify \update{AI}-generated content \update{during the time of development activity}. 
Recent limited-scale empirical work suggests that such comments can indicate generative AI-induced self-admitted technical debt~\cite{mujahid2026genai_satd}.
Figure~\ref{fig:comments-example} shows an example of LLM-referenced comments observed in GitHub repositories, illustrating how developers describe, critique, or justify AI-assisted code within their projects.

This study investigates how developers use, reflect on, and modify \update{AI-assisted} code as noted in code comments in open-source GitHub projects. We focus on developer-authored, self-admitted code comments that explicitly reference \update{AI} use. \update{These comments, along with associated code blocks, provide evidence and insights on how developers are using AI during real-time development. The subsequent commits that change AI-assisted code provide the nature of post-integration modification and adaptation of AI support.}
To guide this investigation, we pose the following research questions:

\smallskip
\noindent
\textbf{RQ1: How do developers integrate \update{AI} into their real-world software development workflows, and what forms of contribution \update{does AI} make?}

\noindent $\boldsymbol{\rightarrow}$
To address this question, we manually annotated 500 unique code comments that explicitly referenced \update{AI} usage and extended the analysis to \update{35,361} comments and code blocks using LLM-based annotation consolidated through the Dawid-Skene (DS) Expectation-Maximization (EM) framework ~\cite{dawid1979maximum}. 
We categorized each instance along two dimensions: \textbf{Task Type}, describing the software engineering activity supported by \update{AI} (e.g., code implementation, enhancement, testing, documentation, or debugging); and \textbf{AI Contribution Type}, representing how the \update{AI} assisted developers (e.g., implementation, knowledge and concept support, or artifact generation).
\update{These two dimensions provide us with insights into the characteristics of AI usage.}
Using both quantitative and qualitative analyses, we find that \update{AI tools and LLMs are }most often used for code implementation and conceptual guidance, positioning them as active collaborators within real-world software development workflows.


\smallskip
\noindent
\textbf{\todo{RQ2: How do developers subsequently adjust, refine, or extend \update{AI}-assisted code after its introduction into projects?}}

\noindent $\boldsymbol{\rightarrow}$
\update{To answer this question, we identified the commits that made changes for the first time to the AI-assisted code block associated with the AI-referenced code comments. We refer these commits as \textbf{first change commits}.}
After applying topic modeling using BERTopic~\cite{grootendorst2022bertopic} on the \textbf{first change commit} messages and semantic grouping, we identified \update{8} types of actions developers mainly applied on the \update{AI}-aided code. The dominant actions were \textit{Refactoring \& Cleanup}, and \textit{Feature Integration \& Extension}. We also observed that a large number of commits involve actions related to \textit{Bug Fixes \& Corrective Changes}.

\smallskip
\noindent
\textbf{RQ3: How has developers’ \update{AI} usage behavior evolved over time?}

\noindent $\boldsymbol{\rightarrow}$
We performed a longitudinal analysis of GitHub \todo{comments} from December 2022 to \update{March 2026} to analyze the evolution of LLM-assisted activities at two levels of granularity from RQ1: a) \textbf{Task Types} and b) \textbf{\update{AI} Contribution Types}.
Over time, \textit{code implementation} remained the dominant activity but gradually declined in relative frequency, while \textit{code enhancement} gained prominence. At the contribution level, \textit{implementation} continued to lead, while \textit{knowledge-seeking} behavior increased most rapidly, suggesting that developers are progressively engaging \update{AI} for conceptual reasoning, design exploration, and informed decision-making rather than solely for code generation.

\noindent
\textbf{Contributions.}
This paper makes the following contributions:
\begin{itemize}  [leftmargin=*]
    \item \textbf{Dataset and Scalable Annotation Scheme.} 
    We curate a dataset of {\update{35,361} GitHub code comments} explicitly referencing AI (e.g., ChatGPT, Copilot, Claude) along with associated code blocks, spanning {December 2022–\update{March 2026}} across {\update{12,944} repositories}. \update{Our dataset also includes 12,778 commits that made changes to these self-admitted AI-assisted code.}
    The dataset is annotated using a \textit{two-stage human-in-the-loop process}, in which 500 manually coded samples guide large-scale LLM-assisted labeling. Annotations from multiple LLMs are consolidated via probabilistic aggregation to ensure reliability and consistency.
    
    \item \textbf{Taxonomy of Developer Tasks and \update{AI} Contributions.}
     We develop a taxonomy of how developers describe and utilize \update{AI} assistance in real-world software development, capturing both \textit{developer task types} (e.g., code implementation, code enhancement, documentation, testing, bug identification \& fixing) and \textit{\update{AI} contribution types} (e.g., implementation, conceptual support, artifact generation).
    
    \item \textbf{Characterization of Post-Integration Developer Actions on \update{AI}-Assisted Code. }
    We link \update{AI}-referenced comments to \update{12,778} \textbf{first-change commits} to capture the \textbf{immediate actions} taken on \update{AI}-assisted code after integration into projects, deriving \textit{seven types of post-integration developer actions} through topic modeling and semantic grouping of commit messages. We further analyze the \textbf{longitudinal evolution} of \update{AI} usage (December 2022–\update{March 2026}), observing a shift from direct code implementation toward increased use of \update{AI} for enhancement, documentation, and conceptual reasoning.

\end{itemize}

\noindent
\textbf{DATA AVAILABILITY.} We provide the annotation guidelines, dataset, and relevant codes to enable replication of our study at~\cite{replication-package}.

\begin{figure*}
    \centering
    \includegraphics[width=\linewidth]{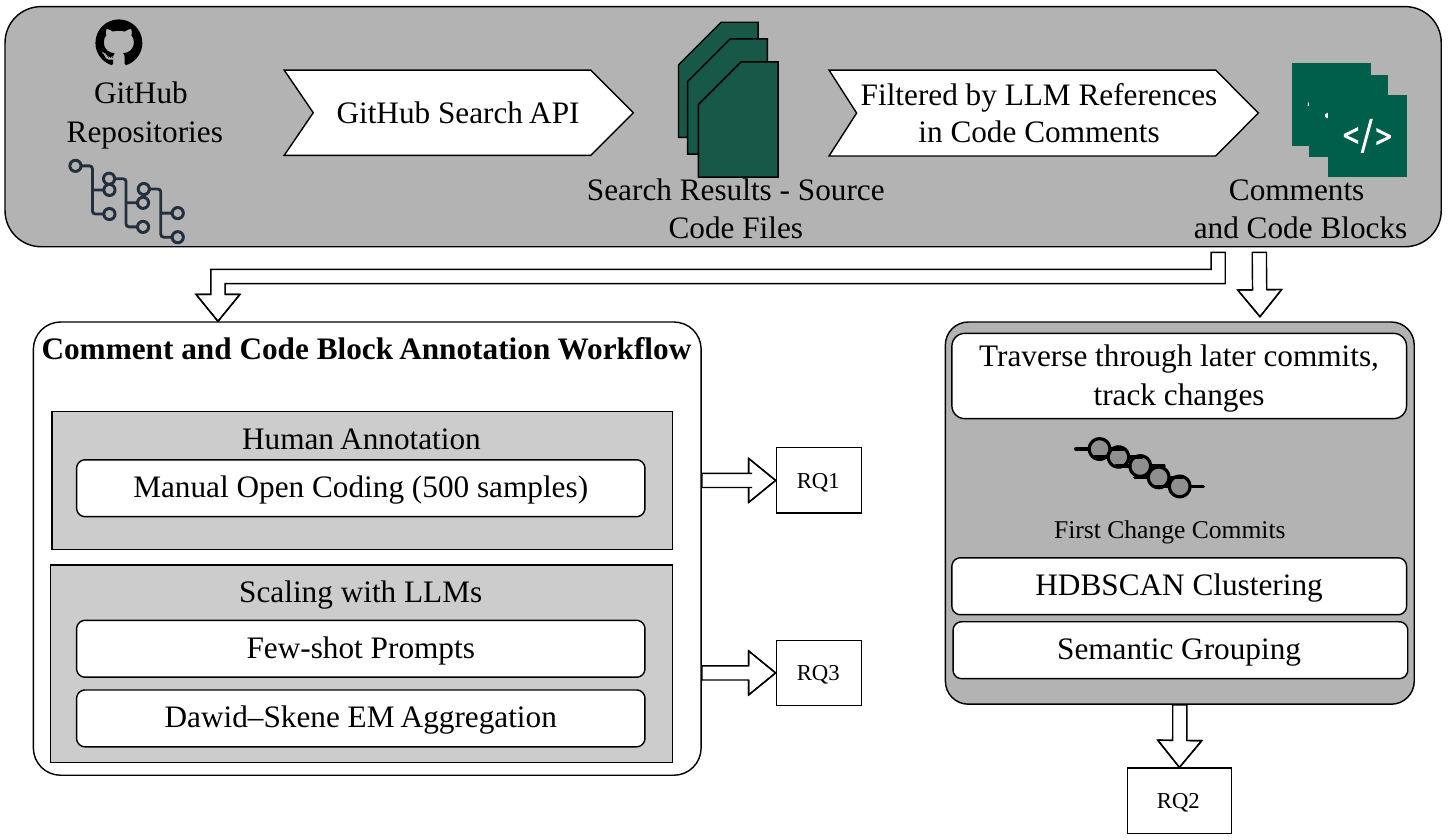}
    \caption{Overview of methodology.}
    \label{fig:pipeline}
\end{figure*}

\section{Methodology}
We adopt a multi-stage, mixed-method research design that combines large-scale data mining, human qualitative coding, probabilistic multi-model annotation, and semantic clustering. Figure \ref{fig:pipeline} provides an overview of the research methodology. We begin by collecting code comments that explicitly reference \update{AI} usage, covering the period from December 2022 (when ChatGPT was first released) to \update{March 2026}. We then extracted the associated code blocks from their introductory commit patches.
The annotation process consists of two phases: (i) manual human annotation and (ii) LLM-based annotation with probabilistic aggregation. RQ1 is examined using this curated and consolidated dataset.
In the subsequent stage, we extract commits those reflecting later modifications to the same code \update{blocks}. In RQ2, we apply HDBSCAN clustering \update{semantic grouping} to these commits to uncover recurrent patterns and underlying motivations driving developers’ modifications and refinements of LLM-assisted code over time. In RQ3, we apply time-series analysis on our derived categories in RQ1 to understand the evolving nature of trends.

\subsection{Data Collection}



We focus on Python and JavaScript, the two most widely used languages in today’s open-source ecosystem~\cite{so2024tech}. Python emerged as the most used language on GitHub in 2024 amid the generative-AI surge~\cite{octoverse2024blog}, while JavaScript has remained a consistent leader in developer activity~\cite{octoverse2023}.

\subsubsection{Extraction of \update{AI}-Referenced Comments \update{and Associated Code Blocks}}

To identify explicit instances where developers referenced the use of LLMs and AI coding assistants as comments in source code, we queried GitHub using the Code Search API with a systematically constructed set of search query templates, as described next. Each template was generated by combining three sets of phrases:

\begin{enumerate}[label=(\Alph*), leftmargin=*]
    \item 16 \todo{keywords identifying different large language models \update{and AI tools}} (e.g., \textit{ChatGPT, GPT, Copilot, Claude, Gemini, Llama, etc}).
    \item 6 action verbs implying generative behavior (e.g., \textit{generated, suggested, created, written, authored,} and \textit{assisted}); and 
    \item 4 connector terms indicating attribution (e.g., \textit{by, from, with}, and \textit{using}).
\end{enumerate}

We applied two complementary query patterns to capture both forward and reversed attributions. The first pattern, (A × B), matched direct expressions such as \textit{“ChatGPT generated”} or \textit{“Claude suggested,”} while the second pattern, (C × B × A), captured reversed constructions such as \textit{“generated by ChatGPT”} or \textit{“written with Copilot”}. The first pattern consists of 96 search queries (16*6) and the second pattern consists of 384 queries (16*6*4) resulting to a total of 480 search queries.
Executing these queries to GitHub Code Search API yielded \update{66,960} matches \update{(43,628 Python and 23,332 JavaScript)}. \update{GitHub Code Search limits results to 1,000 files per query, so retrieval per query is capped at this threshold.}

\update{We first filtered the matches from GitHub search results in two stages. In the first stage, we retained only matches in which the target keyword appeared inside an actual comment span, rather than in code, string literals, identifiers, or other non-comment text, thereby removing false matches from the raw search results. We considered inline comments, block comments, and docstrings, ensuring coverage across language-specific comment syntaxes for both Python and JavaScript. In the second stage, we traced the history of each matched file and inspected its commit patches to locate the earliest commit in which the matched comment text appeared in added lines. The date of this \textbf{introductory commit} was taken as the introduction date, and we retained only those comment records whose introduction date fell within the study window from December 2022 through March 2026. After this filtering, the dataset contained 35,361 comments, corresponding to 26,563 source files from 12,962 unique repositories. \\ We then extracted the associated code block from the introduction commit patch. For each retained record, we located the patch hunk containing the matched comment and selected the smallest code unit still justified by the patch evidence. A hunk was treated as clearly delimiting an \texttt{entity\_block} when the added lines and immediate patch context exposed a self-contained structural unit, such as a function, method, class, test, module-level helper, or similarly bounded region whose beginning and end could be followed directly in the patch. When no complete entity boundary was visible, but the patch still linked the comment to a compact contiguous changed region, we reported that region as a \texttt{local\_code\_span}; in the paper-facing scheme, this category also absorbs cases where the strongest justified unit was the hunk itself. When the introduction commit effectively added an entire file, we retained a \texttt{file\_addition\_block}. In rare cases where the attribution comment could be isolated but no surrounding code region could be justified from the patch, we recorded a \texttt{comment\_only} block. Applying this procedure yielded 35,278 extracted commit-based code blocks: 29,379 \texttt{entity\_block}, 4,374 \texttt{local\_code\_span}, 1,521 \texttt{file\_addition\_block}, and 4 \texttt{comment\_only}, leaving 83 unresolved cases. Of these unresolved cases, 82 were due to commit-fetch failures and 1 was due to missing patch text after the relevant introduction commit had already been identified.}

\update{Finally, we obtained 35,278 (24,882 Python and 10,396 JavaScript) comments with associated code blocks from 26,490 source files from 12,944 repositories.}

\subsubsection{\update{Retrieving First-Change Commits}}
\label{subsubsec:link-comment-to-commit}

\update{To analyze what activities developers performed on AI-attributed code changed after integrating them, we identified \textbf{first-change commit}s for the extracted blocks associated with the comments. For each record, we began from the previously identified introductory commit and reconstructed the extracted block as a line interval in the post-introduction version of the corresponding file. We then followed the history of that same file forward in time and examined later commits in chronological order. For each later commit, we inspected the file patch and updated the tracked interval to account for insertions and deletions that shifted line positions over time. The earliest later commit whose patch overlapped the tracked interval was recorded as the \textbf{first-change commit} to that block.} 

\update{
We successfully retrieved 12,996 \textbf{first-change commit}s and collected commit SHA, author, date, and message. For 22,282 code blocks, the \textbf{first-change commit} could not be retrieved because of insufficient commit history.}

\subsection{Data Annotation and Validation}\label{sec:data-annotation}

\subsubsection{Selection}
A stratified random sample of 500 (balanced by language) comments \update{along with associated code blocks} was selected for manual coding \todo{maintaining the requirements for achieving statistical significance at a 95\% confidence level with a ±5\% margin of error ~\cite{ifeedback_sample_size_calculator,geopoll_sample_size_research}}. Two annotators (with 6+ years of programming experience) independently examined each comment and code block to answer: \textit{“What \update{development} task did \update{AI} assist with, and how?”}. Each \update{instance} was annotated along two dimensions:

\noindent
(1) \textbf{Task Type}, describing the developer activity (e.g., implementation, debugging, documentation, testing, or refactoring); and

\noindent
(2) \textbf{\update{AI} Contribution}, indicating how \update{AI} supported the task (e.g., code generation, suggestion, or auxiliary artifact creation).

\smallskip
\noindent
They were given the following instructions:
\begin{tightquote}
    Please review the annotation instructions carefully. Then, annotate each comment \update{and code block} along two dimensions: \textbf{Task Type} and \textbf{\update{AI }Contribution Type}. Assign multiple codes if needed. If \update{AI} is mentioned but not actually used, mark it as \textit{False Positive} in both dimensions. 
    Use open coding to identify initial patterns, and during axial coding, \textbf{select the single most applicable category} that best represents each comment \update{and code block} to merge related patterns into broader themes, forming the final taxonomies.
\end{tightquote}

We followed an iterative grounded process. The annotators first performed open coding to identify recurring themes, then refined them into a taxonomy of task and contribution categories through axial coding. Disagreements were resolved through several rounds of discussion, resulting in a consistent coding scheme.

During annotation, a substantial number of false positives were detected. We observe instances where \update{AI and} LLM-related terms appeared incidentally in comments (e.g., {someone named Claude commented: ``\texttt{written by Claude Pageauoment}''). 
To maintain precision, these cases were excluded, and additional samples were drawn and annotated until a sufficient volume of valid, \update{AI}-referenced comments was achieved. A total of 173 false positives were removed over 3 rounds of annotation. This process expanded the manually verified set beyond the initial 500 instances. 

After axial coding, we identified six \textbf{Task Types}: \textit{Code Implementation} (362/500), \textit{Code Enhancement} (20/500), \textit{Bug Identification \& Fixing} (26/500), \textit{Testing} (23/500), \textit{Documentation} (16/500), and \textit{Generic Mention and Indeterminate Actions} (97/500); and four \textbf{\update{AI} Contribution Types}: \textit{Implementation} (387/500), \textit{\knowledge{}} (50/500), \textit{Artifact Generation} (15/500), and \textit{Generic Mention and Indeterminate Actions} (48/500). 

While the label \textit{Generic Mention and Indeterminate Actions} appears in both \textbf{Task Type} and \textbf{\update{AI} Contribution Type} categories, it serves different analytical purposes in each dimension. In \textbf{Task Type}, it denotes instances where the developer’s activity could not be identified, whereas in \textbf{\update{AI} Contribution Type}, it indicates that the model’s role was unclear or unspecified. This distinction is further supported by their differing frequencies in the annotated data. 
For example, a comment such as \texttt{``Suggested by CoPilot''}, indicates that a suggestion was taken from LLM, but it does not reveal any specific task category.


We computed inter-annotator agreement on the dataset prior to disagreement resolution.
Given the strong class imbalance in both label spaces, we used Gwet’s AC1~\cite{gwet2014handbook}, which is less sensitive to category-prevalence and marginal-distribution effects under skewed class frequencies~\cite{wongpakaran2013comparison}.
This yielded $AC1 = 0.760$ for \textit{Task type} and $AC1 = 0.657$ for \textit{\update{AI} Contribution type}, corresponding to substantial agreement ($\geq 0.6$ threshold)~\cite{walsh2022assessing}, and indicating a high level of annotation consistency.

\subsection{Scaling Annotation with LLMs}\label{sec:llm-annotation}

Since manually annotating rest of the dataset of \update{35,278} code comments \update{with associated code blocks} is not feasible, we used \textbf{few-shot prompting strategy} integrated with \textbf{Dawid-Skene (DS) Expectation-Maximization (EM)}-based label aggregation framework~\cite{dawid1979maximum,whitehill2009whose}, to extend annotation to the full corpus, as suggested in \textit{LLM-based annotation standardization framework}~\cite{imran2026olaf}. 
Given annotations on the same data by multiple LLMs, the DS-EM method infers the most probable true label by modeling annotator reliability and agreement, allowing multiple LLM outputs to be combined into a single consensus label. 
It has been widely adopted in NLP for combining crowd-sourced or LLM annotations~\cite{he2024if,snow2008cheap,hovy2013mace,gao2024bayesian,yao2024bayesian,ibrahim2025learning}.

\subsubsection{LLM-Based Labeling}
We employed two open-weight models, \textit{gemma-4:31b}~\cite{gemma4_website} and \textit{nemotron-3-super:120b}~\cite{chandiramani2026nemotron}, to enhance robustness and mitigate model-specific bias. 
We used ollama \update{cloud} and set temperature to 0 for both models.
By combining outputs from these distinct architectures, we obtain probabilistically aggregated labels that better capture the consensus between models and reduce variance arising from model-specific behaviors. 

We used two structured prompts to guide the process:
(1) \textbf{Task Type} prompt, which categorized each comment \update{with the associated code block} according to the 6 software development activity identified during axial coding; and
(2) \textbf{\update{AI} Contribution Type} prompt, which identified how \update{AI} contributed based on 4 categories derived from the same coding process. 
If \update{an instance} did not fit any category, we instructed the LLMs to label it as \textit{False Positive}. 
The prompt templates are available in the replication package~\cite{replication-package}.

Each comment \update{and code block} was independently annotated by both LLMs, producing parallel label sets. 
For \textbf{Task Type} annotation, the annotated labels by \textit{gemma-4:31b} and \textit{nemotron-3-super:120b} were same for \update{23,497} instances. For \update{11,781 instances}, the LLMs predicted different labels. 
For \textbf{\update{AI} Contribution Type} annotations, both LLMs predicted the same labels for \update{20,139} instances, and in \update{15,139} cases they predicted different labels. 

These labels were subsequently consolidated using the Dawid–Skene Expectation-Maximization (DS-EM) aggregation procedure~\cite{dawid1979maximum}, which we describe next. 

\subsubsection{Dawid-Skene Expectation-Maximization (DS-EM) Aggregation}
\label{sec:ds-em}

The Dawid-Skene EM algorithm jointly estimates (1) the latent true label for each instance and (2) the reliability of each annotator—in this case, the two LLM classifiers~\cite{dawid1979maximum,whitehill2009whose}. 
Each item $i$ has a latent class $Y_i \in \{1,\dots,K\}$ with prior $\pi_k = P(Y=k)$. 
Observed labels $\ell_i^{(A)}$ and $\ell_i^{(B)}$ follow confusion matrices $C^{(A)}$ and $C^{(B)}$, assuming independence:
$$
P(\ell_i^{(A)}, \ell_i^{(B)} \mid Y_i = k) = C^{(A)}_{k,\ell_i^{(A)}} C^{(B)}_{k,\ell_i^{(B)}}.
$$
The EM procedure alternates between estimating posteriors 
$p_i(k) \propto \pi_k C^{(A)}_{k,\ell_i^{(A)}} C^{(B)}_{k,\ell_i^{(B)}}$ 
and updating priors and confusion matrices with Dirichlet smoothing~\cite{gelman2013bda3}. 
Diagonal priors are biased ($\alpha_{kk} > \alpha_{k\ell}$) to encode higher annotator accuracy.

\update{400 }Gold-standard human annotations from Section \ref{sec:data-annotation} were used to initialize $\pi$ and $C$, anchoring the EM process. 
Training continues until convergence ($\Delta \mathcal{L} < 10^{-6}$) or 100 iterations. 
Hyperparameters: $\alpha_{\text{diag}} = 2.0$, $\alpha_{\text{off}} = 0.5$, $\gamma_\pi = 10^{-3}$. 
Each instance yields a posterior vector $p_i(\cdot)$, hard label $\hat{y}_i = \arg\max_k p_i(k)$, and confidence margin. 

This setup was applied to both \textbf{Task Type} and \textbf{\update{AI} Contribution Type} annotations.

\begin{figure*}[tb]
    \centering
    \includegraphics[width=\linewidth]{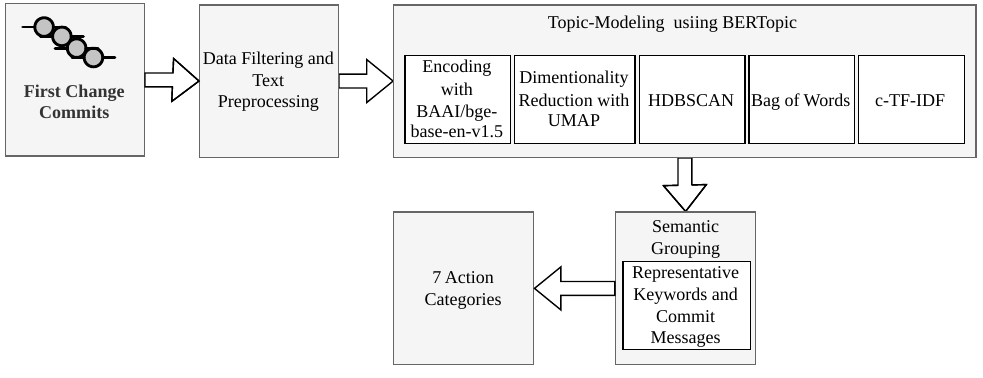}
    \caption{Overview of Topic Modeling and Semantic Grouping.}
    \label{fig:topic-modeling}
\end{figure*}

\subsubsection{Heldout Evaluation}
Out of the 500 manually annotated comments \update{and code blocks}, we stratified-sampled 400 to form a gold set to initialize and anchor DS-EM and reserved the remaining 100 for heldout evaluation. 
We calculated Gwet’s AC1 between the DS-EM outputs and human annotations~\cite{gwet2014handbook}.
On the heldout set, we obtain \textbf{Task Type} AC1 = 0.7005 and \textbf{\update{AI} Contribution Type} AC1 = 0.8128, indicating substantial to near-perfect agreement~\cite{walsh2022assessing}.

\subsubsection{DS-EM Annotation}
DS-EM identified \update{7,697} comments \update{and code blocks} as \textit{False Positives}. 
We excluded these instances from our \update{analysis}.The final aggregated corpus comprised \update{27,581} comments \update{with associated code blocks}, categorized as follows:
\textbf{Task types:} \textit{Code Implementation (\update{18,149}), Code Enhancement (\update{4,039}), Bug Identification \& Fixing (\update{388}), Testing (\update{1,322}), Documentation (\update{1,295}),} and \textit{Generic Mention \& Indeterminate Actions (\update{2,388})}; and 
\textbf{\update{AI} Contribution types:} \textit{Implementation (\update{16,137}), \knowledge{} (\update{2,798}), Artifact Generation (\update{714})}, and \textit{Generic Mention \& Indeterminate Actions (\update{7,932})}. 


\subsubsection{Summary of Comment and \update{Code Block Data}}
Table~\ref{tab:summary} shows the summary of the \update{comment and code block }data and annotation.
We collected \update{26,490} source files containing \update{35,278} LLM-referenced code comments from \update{12,944} repositories. 
Of these, we manually annotated 500 comments. 
The rest were annotated using DS-EM method. 


\begin{table}[tb]
    \footnotesize
    \centering
    \caption{\update{Comment and Code blocks} Summary and Annotation}
    \begin{tabular}{|l|c|c|c|}
    \hline
    & Python & JavaScript & Total \\ \hline
Repositories & \update{9,571} (73.94\%) & \update{3,373} (26.06\%) & \update{12,944}\\ \hline
Files & \update{18,100} (68.33\%) & \update{8,390} (31.67\%) & \update{26,490} \\ \hline
Comments and Code Blocks & \update{24,882} (70.53\%) & \update{10,396} (29.47\%) & \update{35,278} \\ \hline
Manual annotation & 296 (59.2\%) & 204 (40.8\%) & 500 \\ \hline
DS-EM aggregation     & \update{24,586} (70.69\%) & \update{10,192} (29.31\%) & \update{34,778} \\ \hline
    \end{tabular}
    \label{tab:summary}
\end{table}

\update{The collected comment matches and commit data is coming from 12,944 repositories. The star count of these repositories is described in Table~\ref{tab:repo_stars}}. 57 repositories were missing metadata, so we could not collect the star count for them.

\begin{table}[tb]
    \centering
    \caption{\update{Repositories by star count}}
    \begin{tabular}{|l|c|c|c|c|c|c|c|}
        \hline
        Stars & 0 & 1 & 2-4 & 5-9 & 10-19 & 20-49 & 50+\\ \hline
        No of Repositories & 8,349 & 1,690 & 1,088 & 479 & 345 & 307 & 569 \\ \hline
    \end{tabular}
    \label{tab:repo_stars}
\end{table}

\subsection{Topic Modeling and Semantic Grouping of First Change Commit Messages}
\label{sec:clustering}
Figure~\ref{fig:topic-modeling} provides the overview of the topic modeling and semantic grouping procedure.
Out of \update{12,996} collected \textit{first change commits}, we removed \update{2,694} commits linked with a \textit{False Positive} in \textbf{Task Type} or \textbf{Contribution Type} as discussed in Section~\ref{sec:llm-annotation}. \update{After that, we filtered the commit messages with \textit{min\_chars=4} and removed 163 short commit messages}.
After the filtering, we retained \textbf{\update{10,139 first-change commits}} for analysis.

We used BERTopic to identify patterns in developer actions within \textbf{first-change commits}~\cite{grootendorst2022bertopic}. In its default formulation, BERTopic follows a \textit{five-step pipeline}: i) it embeds documents using transformer-based sentence embeddings, ii) reduces embedding dimensionality (e.g., with UMAP~\cite{mcinnes2018umap}), iii) clusters documents using density-based clustering (e.g., HDBSCAN~\cite{mcinnes2017hdbscan}), iv) constructs a cluster-level bag-of-words representation, and v) derives interpretable topic representations using class-based TF-IDF (c-TF-IDF), without requiring a pre-specified number of topics.

Before applying BERTopic,
we performed standard text preprocessing, including lowercasing, and lemmatization to reduce inflectional variation. 
The cleaned commit messages were encoded using dense sentence embeddings generated by \textit{BAAI/bge-base-en-v1.5}, an embedding model designed for retrieval and semantic similarity tasks~\cite{muennighoff2023mteb}. We selected this model since commit messages are typically brief, and semantically sparse. 

As suggested by the authors of BERTopic, we then applied UMAP~\cite{mcinnes2018umap} for dimensionality reduction and HDBSCAN~\cite{mcinnes2017hdbscan} for density-based clustering. 
After doing parameter sweeping~\cite{ruppert2004elements}, we configured UMAP with \textit{n\_neighbors=10, n\_components=10, and min\_dist=0.2}, followed by HDBSCAN with \textit{min\_cluster\_size=10 and min\_samples=5},
enabling the identification of coherent clusters while filtering low-density noise. This pipeline resulted in \update{388} topics containing \update{8,717} commit messages excluding noises. 
We then applied a count-based vectorizer with English stop-word removal and bi-gram and tri-gram features to support class-based TF-IDF (c-TF-IDF) computation, which was used to extract representative phrases for interpreting each topic.

Although these topics capture localized semantic patterns, many reflect closely related developer actions expressed using different surface forms (e.g., \textit{`fix bug'}, \textit{`error fix'}, \textit{`improve error handling'}). To obtain analytically meaningful action categories aligned with developer intent, we grouped semantically similar clusters into higher-level action topics, as fine-grained distinctions in short commit messages are often unstable~\cite{hassan2008automated,tian2012information}. This grouping followed the guidelines in the BERTopic documentation~\cite{grootendorst_bertopic_algorithm}.

We performed the grouping process through semantic grouping, guided by inspection of (i) topic-level keyword representations and (ii) representative commit messages. Each topic contained 6-10 representative keywords. We manually inspected these keywords along with 10 randomly selected commit messages from each 46 topics. We merged topics that described the same underlying activity even when their lexical cues differed. For example, we grouped topics related to error correction, bug fixes (e.g., \textit{`error handling'}, \textit{`bug fix'}, \textit{`py fix'}) under \textit{Bug Fixes \& Corrective Changes}. Similarly, we merged topics capturing additions of new features, UI elements, or functional capabilities under \textit{Feature Development \& Functional Expansion}. In contrast, we grouped topics involving code restructuring, formatting, or removal of unused elements under \textit{Refactoring \& Cleanup}. We utilized thematic analysis to name and group the topics~\cite{maguire2017doing}. Initially, one author coded the initial topics, which were then reviewed and refined by a second author through discussion until thematic saturation was reached~\cite{saunders2018saturation}.
This process resulted in \textbf{\update{eight}} developer action categories: \update{\textit{Feature Integration \& Extension (130 topics, 2,769 commit messages), Refactoring \& Cleanup (59 topics, 1,465 messages), Bug Fixes and Corrective Changes (66 topics, 1,304 messages), Configuration, Dependencies \& Environment Management (20 topics, 810 messages), Documentation (34 topics, 709 commit messages), Testing \& Evaluation (15 topics, 432 messages), Data, Schema \& Pipeline Processing (7 topics, 128 messages), Logging \& Monitoring (4 topics, 72 messages) }. 52 topics containing 1,028 commit messages were miscellaneous updates; those commit messages did not provide enough context to determine the action. For example, ``\texttt{first commit}'', ``\texttt{intermediate commit}'', these commit messages do not clarify the type of action developers performed.}

\subsection{Longitudinal Analysis for Task Type and \update{AI} Contribution Type}\label{sec:method-longitude}

We aggregated monthly annotation counts for both \textbf{Task Type} and \textbf{LLM Contribution Type}
based on each comment's \textbf{introductory commit} timestamp between December 2022 and \update{March 2026}, forming a continuous longitudinal series.  
Categories labeled as \textit{Generic Mention and Indeterminate Actions} were excluded to reduce semantic noise and isolate interpretable actions.  
Each series was then normalized by its monthly total, yielding proportional rather than absolute frequencies to enable cross-category comparison~\cite{quinn2018understanding}.  

To attenuate short-term fluctuations, we applied a three-month rolling mean smoother, providing temporal continuity while preserving structural variation~\cite{box2015time}.  
Sharp spikes and irregularities were further corrected using an Interquartile Range (IQR)-based anomaly cap, clipping values outside the $[Q_1 - 1.5 \times IQR,\, Q_3 + 1.5 \times IQR]$ range~\cite{tukey1977exploratory,chandola2009anomaly}. This adjustment minimized distortion from one-off surges in specific categories.  

From the smoothed and corrected series, we computed descriptive statistics for each label: \textit{mean}, \textit{standard deviation}, \textit{lag-1 autocorrelation} ($\rho_1$), \textit{linear trend slope per month} ($\beta$), and the corresponding \textit{p-value} assessing the statistical significance of $\beta$. These metrics capture the stability, variability, and directionality of LLM activity over time~\cite{box2015time}.
All computations were performed on the corrected normalized data.

\section{Results and Discussion} In this section, we discuss the findings of each research question. 

\subsection{\textbf{RQ1: How do developers integrate \update{AI} into their real-world software development workflows, and what forms of contribution do these models make?}}

To study how developers use \update{AI}, we look at two aspects: what kinds of tasks \update{AI} is used for and how it helps developers during those tasks. Thus, \textbf{RQ1.A} identifies the types of development tasks involving \update{AI}, and \textbf{RQ1.B} examines how \update{AI} contributes to developers’ workflows.

\smallskip
\noindent
\textbf{RQ1.A: \textit{(In-situ Tasks)}. What types of tasks do developers use \update{AI} for?}

\noindent
Our analysis of 500 manually annotated code comments \update{and associated code blocks} identifies six categories of \update{AI}-assisted development activities. This includes a \textit{Generic / Indeterminate} category ({97/500, 19.40\%}) representing cases where LLMs were mentioned without a clear task context. We excluded them from subsequent analysis.
The remaining five task-specific categories are: \textit{Code Implementation} ({362/403, 89.82\%}), \textit{Code Enhancement} ({20/403, 4.96\%}), \textit{Bug Identification \& Fixing} ({26/403, 6.45\%}), \textit{Testing} ({23/403, 5.71\%}), and \textit{Documentation} ({16/403, 3.97\%}).
Table~\ref{tab:rq1a_dev_level} summarizes the frequency and examples of the categories. 

\begin{table*}[!tbh]
\centering
\footnotesize
\caption{Developer-Level Task Distribution of \update{AI} Usage in GitHub}
\label{tab:rq1a_dev_level}
\begin{tabularx}{\textwidth}{|>{\raggedright\arraybackslash}p{0.15\textwidth}
                            |X
                            |X
                            |>{\centering\arraybackslash}p{0.12\textwidth}
                            |>{\centering\arraybackslash}p{0.11\textwidth}|}
\hline
\textbf{Category} &
\textbf{Description} &
\textbf{Example} &
\textbf{Annotated Dataset} &
\textbf{Full Dataset} \\
\hline

Code Implementation &
Generation of functional code for programming tasks, database queries, or other implementation-related work. &
\texttt{"This function was created by Claude AI."} &
362 (89.82\%) &
18,149 (72.04\%) \\
\hline

Code Enhancement &
Improvement of existing code, including readability, performance, maintainability, or error handling. &
\texttt{"This function was generated using ChatGPT with the prompt: Improve [...]"} &
20 (4.96\%) &
4,039 (16.03\%) \\
\hline

Testing &
Support for test-case generation, testing strategy design, or validation activities. &
\texttt{"Description: this file contains test cases generated by Copilot."} &
23 (5.71\%) &
\update{1,322} (5.25\%) \\
\hline

Documentation &
Generation or refinement of comments, docstrings, technical documentation, or explanatory text. &
\texttt{"Generated by ChatGPT to document the function behavior."} &
16 (3.97\%) &
\update{1,295} (5.14\%) \\
\hline

Bug Identification \& Fixing &
Identification, diagnosis, or correction of bugs, defects, runtime errors, or faulty behavior. &
\texttt{"Suggested by ChatGPT - Fixes PyCharm backend crash"} &
26 (6.45\%) &
\update{388} (1.54\%) \\
\hline

\textbf{Total} &
&
&
\textbf{403} &
\textbf{25,193} \\
\hline

\multicolumn{5}{l}{$\ddag$ \textit{Generic mentions and indeterminate actions are excluded.}}
\end{tabularx}
\end{table*}

\textit{Code Implementation} involves developers using \update{AI tools and LLMs} to generate functional code that becomes part of production repositories. 
For example, comments like “\texttt{This function was created by Claude AI.}” illustrate model-driven code generation embedded within active projects. 
\textit{Code Enhancement} represents developers using \update{AI} to refactor, improve, or optimize existing code, for instance, “\texttt{This function was generated using ChatGPT with the prompt: ‘Improve the delete\_task function with better error handling and improved readability.’}”, highlighting how \update{AI} assists in iterative improvement during development.

Developers also used Generative AI for bug fixing, code quality assurance, and documentation tasks. \textit{Bug Identification \& Fixing} includes AI-suggested fixes and patches such as “\texttt{Suggested by ChatGPT – Fixes PyCharm backend crash.},” demonstrating their role in detecting and resolving issues within active codebases. 
\textit{Testing} involves model-generated or suggested test cases integrated directly into testing pipelines. \textit{Documentation} encompasses \update{AI}-generated docstrings, inline comments, and file-level descriptions, e.g., “\texttt{Description: this file contains test cases generated by Copilot.}” Finally, \textit{Generic Mentions \& Indeterminate Actions}  capture cases where \update{AI} usage is acknowledged without additional context or specification.

Extending this taxonomy to the full dataset of \update{27,581} instances using DS-EM framework, we observed that a substantial portion of the comments \update{and code blocks} were categorized as \textit{Generic Mentions and Indeterminate Actions} (\update{2,388, 8.66\%}). After excluding these instances, among the remaining \update{25,193} task-specific cases,
\update{\textit{Code Implementation} (18,149, 72.04\%) remains the most dominant activity. 
However, the relative presence of \textit{Code Enhancement} increases from 4.96\% in the human annotation dataset to 16.03\% (4,039) in the full dataset, indicating that developers are not only using AI for code generation but also for improving performance, correctness, and efficiency of existing code. Although \textit{Bug Identification and Fixing} was found in 6.45\% of cases in the human annotation, in the full dataset, it was found only in 1.54\% of cases, indicating that developers are not often seeking help from AI for identifying and fixing bugs.}


\smallskip
\noindent
\textbf{RQ1.B: \textit{(Forms of Assistance}). How does \update{AI} assist developers in performing these tasks in practice?}

Building on RQ1.A where we looked for what types of development tasks developers use \update{AI} for, in this RQ, we investigate how \update{AI} provides support during these development tasks. 

The manual coding of 500 code comments \update{and code blocks} revealed three primary forms of assistance: \textit{Implementation} (387/452; 85.62\%), \textit{\knowledge{}}  (50/452; 11.06\%), and \textit{Artifact Generation}  (15/452; 3.32\%). The rest were \textit{Generic Mention and Indeterminate Actions} (48).
Similar to  RQ1.A, we extended this taxonomy for the full dataset of \update{27,581} comments \update{code blocks}. \update{7,932, (28.79\%)} comments were labeled as \textit{Generic Mention and Indeterminate Actions}.
Table~\ref{tab:contribution-type} shows the distribution of \update{AI} contribution types, along with examples, in both the manually annotated and the full dataset of \update{19,649} comments \update{and code blocks}.

\begin{table*}[!tbh]
\centering
\footnotesize
\caption{\update{Distribution of AI Contribution Types in GitHub}}
\label{tab:contribution-type}

\begin{tabularx}{\textwidth}{|>{\raggedright\arraybackslash}p{0.13\textwidth}
                            |X
                            |X
                            |>{\centering\arraybackslash}p{0.12\textwidth}
                            |>{\centering\arraybackslash}p{0.11\textwidth}|}
\hline
\textbf{Category} &
\textbf{Description} &
\textbf{Example} &
\textbf{Annotated Dataset} &
\textbf{Full Dataset} \\
\hline

Implement-ation &
\update{AI directly contributes to implementation by generating, completing, or modifying source code.} &
\texttt{"Validation fully written by Copilot based on error state names"} &
387 (85.62\%) &
16,137 (82.13\%) \\
\hline

\knowledge{} &
\update{AI provides conceptual guidance, suggestions, or domain knowledge in response to a developer query.} &
\texttt{"Using bcrypt for password hashing (ChatGPT suggested secure hashing practices"} &
50 (11.06\%) &
2,798 (14.76\%) \\
\hline

Artifact Generation &
\update{AI generates non-code artifacts, such as documentation, text, images, configuration files, or structured lists.} &
\texttt{"This is a ChatGPT generated list of devices"} &
15 (3.32\%) &
714 (3.63\%) \\
\hline

\textbf{Total} &
&
&
\textbf{452} &
\textbf{19,649} \\
\hline

\multicolumn{5}{l}{$\ddag$ \textit{Generic mentions and indeterminate actions are excluded.}}
\end{tabularx}
\end{table*}

The findings indicate that the vast majority of interactions involve \textit{Implementation}, where developers use \update{AI} to produce executable code or complete functional logic that becomes part of a repository. In this mode, \update{AI tools and LLMs} act as direct contributors rather than advisory tools. Comments such as \texttt{"Validation fully written by Copilot based on error state name"} or \texttt{"Function generated by ChatGPT for sorting users by last activity"} exemplify this behavior.

A subset of \update{contributions} captures \textit{\knowledge{}} interactions. In the full dataset, 
{\update{14.76}\%} comments \update{and code blocks} indicate developers seeking knowledge, idea or suggestions from \update{AI}. 
A closer look on the 50 manually annotated \update{instances} shows diverse areas where developers leveraged \update{AI} to obtain conceptual or technical guidance. 
Developers treated \update{AI} as an on-demand advisor to explore design decisions, clarify options, or refine implementation strategies. For instance, \texttt{"Using bcrypt for password hashing (ChatGPT suggested secure hashing practices)"} illustrates how \update{AI} acted as \textit{conceptual partners}, offering reasoning and best-practice advice.
In {12 out of 50} cases, developers sought assistance with \textit{algorithm selection or data structure design}, aiming to identify efficient computational strategies or representation techniques. Another {8 comments} focused on \textit{performance optimization and debugging}, where \update{AI} helped diagnose runtime issues or improve execution. Additionally, {6 comments} dealt with \textit{syntax, framework functions, or API behavior}, demonstrating how developers used \update{AI} to clarify environment-specific technical details.
The remaining interactions involved broader exploratory guidance, reflecting developers’ use of \update{AI} for high-level reasoning and decision support.

The third most prevalent type of contribution is \textit{Artifact Generation}. In these cases, \update{AI} generates supplementary artifacts that go beyond code, such as documentation, test descriptions, or artifacts supporting the development process. Comments like "\texttt{This is a ChatGPT-generated list of device}" represent this category. \update{We found \textit{Artifact Generation} in 714 (3.63\%) instances.}

\smallskip
\noindent
\underline{\textbf{RQ1 Summary}.}
Developers use \update{AI} as active collaborators across multiple stages of software development. Most interactions involve \textit{Code Implementation}, where models generate production-level code, confirming their role as direct contributors. \textit{Code Enhancement} and \textit{Documentation} reflect developers’ use of \update{AI} for refining existing code and producing descriptive artifacts. \textit{\knowledge{}} interactions show that developers also treat \update{AI} as advisory systems, consulting them for design decisions, debugging, and best practices. Although \textit{Artifact Generation} occurs less frequently, it highlights \update{AIs}’ role in creating supporting materials such as `\textit{configs}' and test templates.

\subsection{\textbf{RQ2: How do developers subsequently adjust, refine, or extend \update{AI}-assisted code after its introduction into projects?}} \label{sec:rq2b}

\begin{table}[tb]
\centering
\small
\caption{Modification Activities Identified from First Change Commit Messages}
\label{tab:rq2b_clusters}
\begin{tabular}{|l|c|p{0.4\textwidth}|} \hline

\textbf{Modification Intent} & \textbf{$N$} & \textbf{Example Commit Message} \\ \hline

\multirow{3}{15em}{Feature Integration \& Extension} & 2,769 & \texttt{"feat: add chat management functionality with MongoDB integration"} \\ \hline

\multirow{1}{15em}{Refactoring \& Cleanup} & 1,465 & \texttt{"PEP8 \& Removed Unused Imports"} \\ \hline

\multirow{3}{15em}{Bug Fixes \& Corrective Changes}  & 1,304 & \texttt{"Fix: Improve SOAP error handling and HTML detection for malformed responses"} \\ \hline

\multirow{2}{15em}{Configuration, Dependency \& Environment Management} 
& 810 & \texttt{"Merge branch `lock\_capsule\_feature' into dev"} \\ \hline
\multirow{2}{15em}{Documentation} & 709 & \texttt{"add types and part of a docstring"} \\ \hline

\multirow{1}{15em}{Testing \& Evaluation}  & 432 & \texttt{"Add tests for multiturn"} \\ \hline

\multirow{2}{15em}{Data, Schema \& Pipeline Processing}  & 128 & \texttt{"need to use OCR instead of pdf plumber for text extraction"}  \\ \hline

\multirow{2}{15em}{Logging \& Monitoring} & 72 & \texttt{"Add logging. Only traverse data directory"} \\ \hline

\end{tabular}
\end{table}

Table~\ref{tab:rq2b_clusters} summarizes the types, frequencies, and representative examples of developer \textit{actions} used to adjust, refine, or extend \update{AI}-assisted code after its integration into real-world GitHub projects. 

\update{
The most frequent action observed in our analysis is \textit{Feature Integration \& Extension (2,769)}. Examples such as \texttt{"feat: add chat management functionality with MongoDB integration"}, \texttt{"update frontend code, for upload images and pagination"}, and \texttt{"feat: enhance chat endpoint to accept both `message' and `prompt' fields for improved flexibility"} show that, after integrating AI-assisted code, subsequent commits commonly extend functionality and introduce new features.}

\update{
The second most prevalent action observed in our analysis is \textit{Refactoring \& Cleanup (1,465)}. Commit messages in this category describe structural refinement, reformatting, code style fixes, and general code cleanup. For example, messages such as \texttt{"remove redundant my\_dx.sample"} and \texttt{"Clean up unused code."} explicitly indicate the removal of redundant or unused code. We also observe commits describing the removal of unnecessary dependencies (e.g., \texttt{"remove unused dependencies"}) and adjustments to functional requirements (e.g., \texttt{"removed username requirements"}). 
Other commit messages, such as \texttt{"refactor: update import organization setting and add test endpoint"}, \texttt{"refactor: small change"}, and \texttt{"Refactor backend state to use singleton instantiation pattern; add support for live song subtitles"}, reflect modifications to AI-assisted code aimed at adjusting functionality, introducing minor changes, or improving design patterns for better usability. Overall, \textit{Refactoring and Cleanup} commit messages indicate that AI-assisted code often undergoes immediate modification and improvement after integration.
}

\update{
\textit{Bug Fixes \& Corrective Changes (1,304)} are commonly observed after the integration of AI-assisted code. For instance, the commit message \texttt{"Authorization issue solved, error handling in progress"} reflects post-integration resolution of an authorization issue. Other messages indicate the identification of bugs (e.g., \texttt{"Handle LabelGraphics bug in dot-gml script"}) and their subsequent correction (e.g., \texttt{"fixed a bug in the onmessage function"}, \texttt{"Fix: Improve SOAP error handling and HTML detection for malformed responses"}), suggesting that AI-assisted code often requires corrective maintenance.
}

\update{
A substantial portion of commit messages falls under \textit{Configuration, Dependency \& Environment Management (810)}. Commits referencing merge conflict resolution (e.g., \texttt{"resolve merge conflict"}) or pull request integration (e.g., \texttt{"Merge pull request \#45 from apfox500/profile-messaging.Profile messaging"}) indicate that AI-assisted code has been successfully incorporated, after which the immediate follow-up actions relate to project coordination rather than code modification.
}

\update{
Commit messages also frequently relate to \textit{Documentation (709)}. These messages indicate updates to README files and the addition of documentation after AI-assisted code is introduced. In addition, a notable number of commits involve \textit{Testing \& Evaluation (432)}. Messages such as \texttt{"added additional testing to improve coverage of all functions"} and \texttt{"Docs: Add manual testing and user story testing"} suggest that additional testing activities commonly follow the introduction of AI-assisted code. We also observed a small number of commit messages related to \textit{Data, Schema \& Pipeline Processing (128)}, for example \texttt{"need to use OCR instead of pdf plumber for text extraction"} and \textit{Logging \& Monitoring (72)}, for example \texttt{"Add prefix cache hit rate to metrics"}.
}

\smallskip
\noindent
\underline{\textbf{RQ2 Summary}.}
Our analysis of \textbf{first-change commit} messages shows a clear split in post-integration actions on \update{AI}-assisted code. A majority of the observed commits focus on modification, refinement, and documentation activities, as reflected in the \textit{Refactoring \& Cleanup}, \textit{Bug Fixes \& Corrective Changes}, \textit{Testing \& Evaluation}, and \textit{Documentation} categories. The remaining commits primarily reflect successful integration and continuation of development, captured by \textit{Feature Integration \& Extension} and \textit{Configuration, Dependency \& Environment Management}. Together, these patterns indicate that while \update{AI}-assisted code is frequently integrated into projects, it commonly undergoes substantial post-integration refinement.

\subsection{\textbf{RQ3: How has developers’ \update{AI} usage behavior evolved over time?}}

\begin{figure}[tb]
    \centering
    \includegraphics[width=1\textwidth]{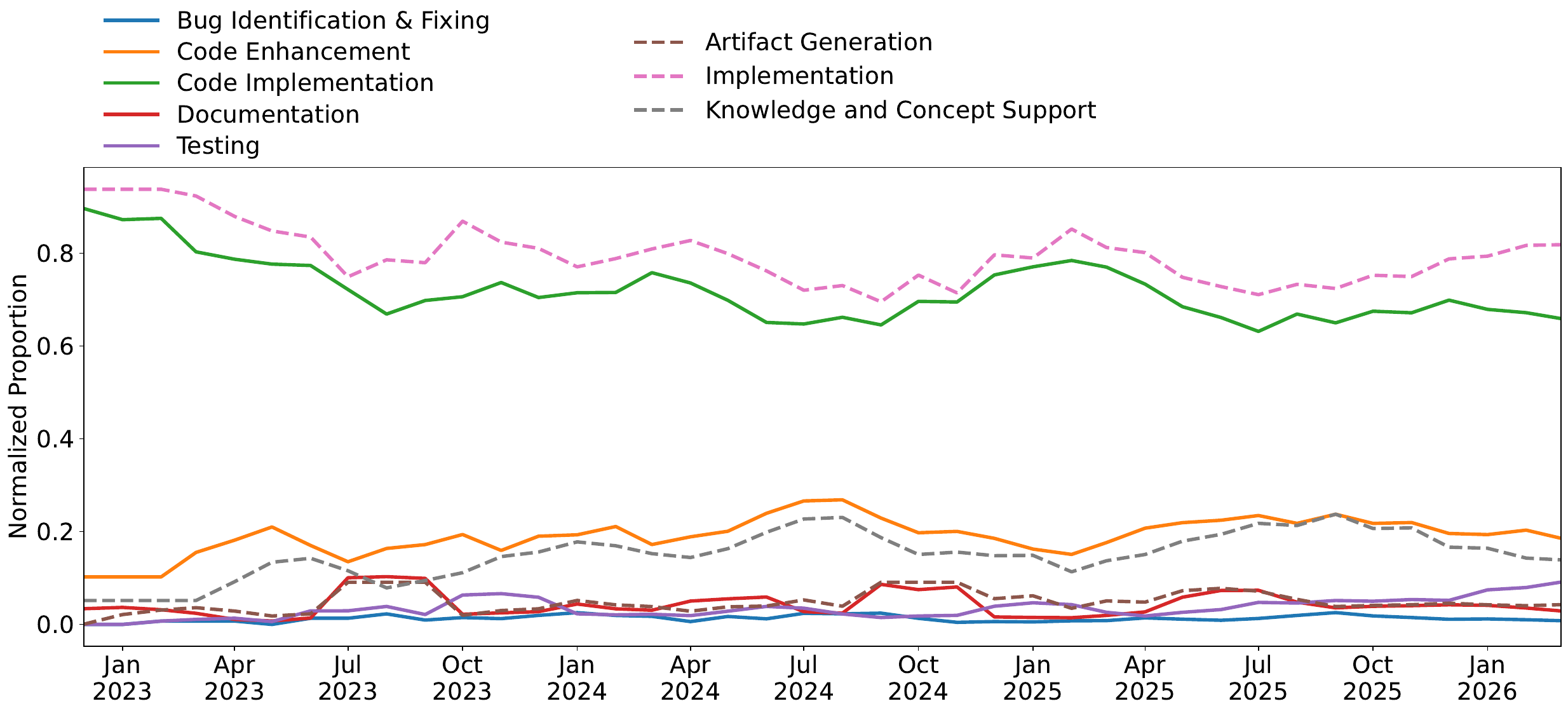}
    \caption{Temporal Evolution of AI-Assisted Development Task Types and AI Contribution Types}
    \label{fig:time-series-1}
\end{figure}

As mentioned in Section~\ref{sec:method-longitude}, to answer this RQ, we conducted a longitudinal analysis of \update{AI}-referenced code comment data from December 2022 and March 2026. 
The distribution of \update{AI}-referenced comments increased substantially over time. Only 0.31\% of records occurred in 2022, followed by 7.80\% in 2023, 19.78\% in 2024, 52.41\% in 2025, and 19.69\% in 2026. This pattern indicates that \update{AI}-referenced commenting activity was minimal in 2022, expanded rapidly during 2023 and 2024, peaked in 2025, and remained substantial in early 2026.

In the following, we used the terminology provided by CDC for trend analysis~\cite{CDC_StatisticalSignificance_2024}: ``\textit{terms such as “stable,” “no clear trend,” and “did not change significantly” indicate that the slope of the trend line was not significantly different from zero. Terms such as “increase” and “decrease” indicate that a significant trend was found.}"

To quantify temporal dynamics, we constructed monthly time series for each category. For each month, we computed the normalized share of each category relative to total activity after removing False Positive and Generic/Indeterminate labels. The series was smoothed using a three-month rolling mean to reduce short-term fluctuations while preserving temporal variation, and extreme values were corrected using an interquartile range-based clipping method.

We then calculated descriptive statistics including mean ($\mu$), standard deviation ($\sigma$), and lag-1 autocorrelation ($\rho_1$). To assess trends ($\beta$), we fitted a linear model where the category share is regressed on time, and the slope represents the monthly rate of change. Statistical significance ($p$) of the slope was evaluated using Newey-West~\cite{newey1986simple} adjusted standard errors to account for autocorrelation.






\begin{table}[tb]
\centering
\caption{Longitudinal statistics of \update{AI} usage by task and contribution type. 
$\mu$ is the mean proportion, $\sigma$ the standard deviation, $\rho_1$ the lag-1 autocorrelation, $\beta$ the monthly trend slope, and $p$ the Newey-West corrected significance value. $^{*} p < 0.05$, $^{**} p < 0.01$, $^{***} p < 0.001$.}
\label{tab:longitudinal-analysis}
\begin{tabular}{|l|c|c|c|c|c|}
\hline
\textbf{Category} & $\boldsymbol{\mu}$ & $\boldsymbol{\sigma}$ & $\boldsymbol{\rho_1}$ & $\boldsymbol{\beta}$ & $\boldsymbol{p}$ \\
\hline

\multicolumn{6}{|l|}{\textbf{Task Type}} \\
\hline

Code Implementation & 0.739 & 0.047 & 0.712 & -0.00080 & 0.449 \\
Code Enhancement & 0.155 & 0.030 & 0.818 & +0.00113 & $0.049^{*}$ \\
Documentation & 0.060 & 0.036 & 0.515 & -0.00151 & $0.008^{**}$ \\
Testing & 0.032 & 0.019 & 0.796 & +0.00097 & $0.004^{**}$ \\
Bug Identification \& Fixing & 0.014 & 0.006 & 0.554 & +0.00021 & $0.025^{*}$ \\
\hline

\multicolumn{6}{|l|}{\textbf{\update{AI} Contribution Type}} \\
\hline

Implementation & 0.814 & 0.044 & 0.631 & -0.00004 & 0.970 \\
Knowledge \& Concept Support & 0.135 & 0.040 & 0.899 & +0.00185 & $0.025^{*}$ \\
Artifact Generation & 0.050 & 0.040 & 0.597 & -0.00181 & $0.006^{**}$ \\
\hline
\end{tabular}
\end{table}

Table~\ref{tab:longitudinal-analysis} summarizes these results. Across task types, \textit{Code Implementation} remains the dominant activity ($\mu = 0.739$), but does not exhibit a statistically significant trend over time ($\beta = -0.00080$, $p = 0.449$). This suggests that implementation remains the primary form of \update{AI}-referenced activity, even as other task types gain visibility. In contrast, \textit{Code Enhancement} shows a significant increase ($\beta = +0.00113$, $p = 0.049$), suggesting that developers increasingly use LLMs for refining and improving existing code rather than generating it from scratch. Similarly, \textit{Testing} demonstrates a significant positive trend ($\beta = +0.00097$, $p = 0.004$), indicating growing reliance on \update{AI} for validation and quality assurance tasks.

\textit{Documentation} shows a statistically significant decrease ($\beta = -0.00151$, $p = 0.008$), suggesting that documentation-related \update{AI} references became less prominent relative to other categories over time. \textit{Bug Identification \& Fixing} also shows a statistically significant positive trend ($\beta = +0.00021$, $p = 0.025$), suggesting that while debugging remains a smaller category overall ($\mu = 0.014$), its relative presence increased over time.

For \update{AI} contribution types, \textit{Implementation} is the most prevalent category ($\mu = 0.814$), but shows no statistically significant trend ($\beta = -0.00004$, $p = 0.970$). This indicates that direct implementation remains dominant but relatively stable over time. In contrast, \textit{Knowledge \& Concept Support} exhibits a significant increase ($\beta = +0.00185$, $p = 0.025$), reflecting a transition toward using \update{AI} as cognitive assistants for explanation, reasoning, and conceptual guidance. \textit{Artifact Generation} shows a statistically significant decrease ($\beta = -0.00181$, $p = 0.006$), indicating that this contribution type became less prominent relative to implementation and conceptual-support activities.

Figure~\ref{fig:time-series-1} further illustrates these dynamics. The temporal trajectories reveal moderate to high persistence across categories, as reflected by the lag-1 autocorrelation values. The rise of enhancement, testing, bug identification and fixing, and conceptual support suggests a gradual redistribution of \update{AI} involvement beyond direct implementation alone.

\underline{\textbf{RQ3 Summary.}}
Overall, the results indicate a clear evolution in developers’ \update{AI} usage behavior. Direct implementation remains the dominant form of \update{AI}-referenced activity, but its relative prevalence is stable rather than significantly increasing or decreasing. Over time, usage diversified toward refinement, testing, debugging, and cognitively oriented support. This shift suggests that \update{AI} is increasingly integrated not only as tools for execution, but as collaborators supporting reasoning, refinement, and software quality processes.


\section{\todo{Discussion and Implications}}

Our empirical analysis clarifies how \update{AI}-assisted development operates in practice. The findings point to several implications for knowledge management, productivity measurement, and long-run adoption dynamics.

\smallskip
\noindent
{\bf Knowledge Externalization and Organizational Memory.}
The increasing use of \update{AI} for knowledge and concept support (RQ1) indicates that developers rely on them to bridge knowledge gaps during task execution, such as understanding APIs, clarifying design choices, or reasoning about implementation strategies.
Evidence from industry observation also indicate that practitioners use conversational LLMs for guidance and learning rather than expecting ready-to-integrate artifacts~\cite{khojah2024beyond}. 

Applied to \update{AI}-assisted development, this has practical risk for organizational learning and continuity~\cite{ackerman1998considering}. 
Research on developer–assistant interaction found that preserving decision rationale and traceability from conversational assistance requires explicit persistence mechanisms and does not occur automatically~\cite{contreras2024conversational}.
Design and implementation reasoning that occurs during LLM conversations will not be visible to later contributors unless it is explicitly recorded. As a result, project history alone may be insufficient to reconstruct decision rationale.

\noindent
{\em Implications.} Teams should ensure that when \update{AI} assistance materially influences a design or implementation decision, a brief rationale is recorded in existing artifacts such as pull requests, issues, or design notes. This relies on established documentation practices and directly addresses the traceability gap identified in prior work.

\smallskip
\noindent
\textbf{The Overhead Cost of Code Integration.}
The substantial post-integration modification effort observed (RQ2) raises questions about how productivity gains from \update{AI} assistance should be measured~\cite{weisz2025examining,mohamed2025impact}. When \update{AI}-assisted code regularly requires refactoring, testing, and debugging after introduction, traditional metrics, such as lines of code generated or time to first implementation may misrepresent actual efficiency improvements.

Total cost of ownership (TCO)~\cite{mieritz2005defining} frameworks in software engineering account for implementation, integration, maintenance, and operational costs beyond initial acquisition. 
The modification patterns we observed suggest that \update{AI} may provide value through different mechanisms than initially assumed, redistributing effort from initial scaffolding to integration and refinement work. This is also observed by Mujahid et al.~\cite{mujahid2026genai_satd}. Measuring such benefits requires tracking developer effort across the entire lifecycle, not just initial code creation.

\noindent
{\em Implications.} 
Researchers and organizations should adopt lifecycle metrics that encompass generation, integration, testing, and maintenance efforts. Evaluation of LLMs can incorporate project-based metrics and assess how generated code integrates into real software engineering workflows, rather than relying solely on isolated benchmark performance. Time and quality tracking across these phases will yield a more accurate picture of cost-benefit trade-offs for \update{AI} utility.

\smallskip
\noindent
{\bf Temporal Maturation of \update{AI}-Assisted Development Practices.}
The longitudinal analysis (RQ3) shows that \update{AI} adoption is accompanied by qualitative shifts in use. The increase in \textit{Knowledge \& Concept Support} and \textit{Code Enhancement} alongside stable \textit{Code Implementation} suggests that developers are moving from exploratory use to more differentiated and deliberate application of \update{AI}. This pattern aligns with the diffusion of innovation theory~\cite{rogers2014diffusion}: developers first adopt \update{AI} for well-defined, low-risk tasks (code scaffolding) before expanding to tasks requiring judgment and discretion (architecture decisions, optimization strategies).

\noindent
{\em Implications.} \update{AI}-assisted development should be understood as a multi-stage practice rather than a single adoption event. Evaluations and tool designs that focus primarily on code generation risk, overlooking where ongoing maturation is occurring, namely in conceptual support and enhancement-oriented use.

\section{Related Work}
We divide the related work into two parts: (1) code comments and commit messages to infer developer intent and activities, and (2) how LLMs are used in software development.

\smallskip
\noindent
\textbf{Code Comments and Commit Messages.}
Textual artifacts such as code comments and commit messages have long been used to understand developer intent and maintenance behavior~\cite{hindle2008large,steidl2013quality,kagdi2007survey,mu2023developer,codabux2024teaching,tan2015code}. Rani et al. reviewed comment quality, showing that comments capture rationale, design trade-offs, and cognitive processes~\cite{rani2023decade}. Hu et al. found that developers value concise, contextually accurate automated comments~\cite{hu2022practitioners}. Hindle et al. classified commit messages into corrective, adaptive, and perfective maintenance categories, showing that message text reflects developer intent and task type~\cite{hindle2009automated}. \todo{Ferreira et al. worked on characterizing github commits and compared them with commit size~\cite{ferreira2022characterizing}. Tian et al. analyzed attributes of good commit messages~\cite{tian2022makes}. Yamauchi et al. used clustering techniques to understand developer intentions from commit messages~\cite{yamauchi2014clustering}.}
Xue et al. observed that LLMs can produce commit messages comparable to human-written ones, highlighting AI’s growing role in developer communication and documentation practices~\cite{xue2024automated}, while
Katzy et al. investigated how multilingual code comment generation by LLMs introduces unique errors and question the reliability of automatic metrics for evaluating AI-generated comments~\cite{katzy2025Qualitative}. Li et al. found that comments presents in AI-generated code increased programmers’ adoption, regardless of expertise \cite{li2026comments}.



\smallskip
\noindent
\textbf{\update{AI} Usage in Software Development.}
Generative AI and coding assistants such as Copilot, ChatGPT, Cursor, and Claude have affected software engineering practices significantly. Fan et al. surveyed large language models in software engineering and described open problems~\cite{fan2023large}.
Barke et al. investigated Copilot usage in real-time editing to characterize collaborative code completion~\cite{barke2023grounded};
Du et al. and Jin et al. evaluated model accuracy and usability in code-generation tasks~\cite{du2024evaluating,jin2024canchatgpt}; 
and Guo et al. examined LLMs' ability to refine or repair code automatically~\cite{guo2024exploring}. The AIDev dataset enables large-scale empirical analysis of AI coding agents in real-world GitHub pull requests~\cite{li2025aidev}.
While these studies demonstrated productivity gains, they captured only short-term and synthetic development scenarios.
Mining-based studies shifted toward understanding real-world LLM usage. 
Grewal et al. analyzed ChatGPT-generated code within GitHub repositories and found that developers adopt and integrate AI-produced snippets into projects~\cite{grewal2024analyzing}. 
Hao et al. studied 580 ChatGPT conversations shared through pull requests and issues, identifying sixteen inquiry types such as debugging, testing, and documentation~\cite{hao2024sharedconversations}. 
Mujahid et al. analyzed 81 code comments which referenced generative AI as well as self-admitted technical debt~\cite{mujahid2026genai_satd}.
Mohamed et al. and Sagdic et al. analyzed the DevGPT dataset and found that developers use LLMs for programming guidance, framework clarification, and explanation of APIs~\cite{mohamed2024chatting,sagdic2024taxonomy}. 
At a broader organizational level, Mozannar et al. and Murali et al. observed that AI-assisted programming alters productivity, workload distribution, and cognitive effort in industrial teams~\cite{mozannar2024reading,murali2024ai}.
Aguiar et al. analyzed practitioner conversations with ChatGPT across multiple programming languages, showing how developers use LLMs for cross-language comprehension, translation, and problem solving in real software development tasks~\cite{aguiar2024multi}. 
Previous works broadly explored the role of large language models in software engineering, highlighting their potential to enhance developer productivity and support tasks such as code generation and comprehension, while also identifying challenges related to reliability, maintainability, human oversight, and long-term risks in software development contexts~\cite{belzner2023large,ozkaya2023application,moroz2022potential,gao2025current,abdelsalam2026humansLLMs,huang2026issuecommitllm,khatib2026assertflip,gu2026semanticrepairllm}. 
While prior studies have examined \update{AI} usage in software development, limited attention has been given to code comments. 
We build on this work by analyzing comment- and commit-level data to examine how developers incorporate and refine \update{AI}-generated code in open-source projects.

\section{Threats to Validity}

Our study is subject to several important validity threats, which we describe below:

\smallskip
\noindent{\textbf{Construct validity.}}
Construct validity concerns the extent to which the methods and measures used in the study accurately capture the intended research constructs. Our detection of \update{AI}-related comments relied on keyword-based queries, which may introduce false positives (e.g., unrelated names like “Claude”). We mitigated this by manually validating a random subset for precision. Annotation bias was reduced through independent labeling by multiple annotators, iterative guideline refinement, and inter-annotator agreement checks. We considered comments \update{and code blocks} to classify developer tasks, though \update{AI}-referenced comments express the developer's intent, this raises a possibility of misinterpreting the actual task. \update{Our captured code blocks might include non AI-assisted code; our blocks can also fail to capture the complete code that was assisted by AI.}

\smallskip
\noindent{\textbf{Internal validity.}}
Internal validity refers to whether the observed relationships genuinely reflect causal or meaningful associations, rather than being influenced by confounding factors or methodological artifacts.
The linkage between \update{AI}-referenced comments and commits assumes repository histories preserve temporal order and authorship. We excluded commits with anomalous timestamps or inconsistent metadata. Because commits occur at the file level and may include unrelated changes, we limited analysis to the commit where the \update{AI}-referenced comment first appeared and the earliest subsequent commit modifying the same file. 

\smallskip
\noindent{\textbf{External validity.}}
External validity concerns how well our findings generalize beyond the specific dataset, repositories, or environments analyzed.
Our dataset focuses on public GitHub repositories in Python and JavaScript, which may not represent other ecosystems or closed-source projects.
Moreover, it captures only self-admitted \update{AI} usage, explicit mentions of tools like ChatGPT or Copilot, thus underrepresenting silent or unacknowledged use.
As \update{AI}-assisted development practices evolve rapidly, our findings should be viewed as representative of current open-source trends rather than the full spectrum of developer behavior.

\section{Conclusion and Future Work}

We analyzed 35,361 AI-referenced code comments and associated code blocks added in GitHub between December 2022 and March 2026 and examined their post-integration trajectories through 12,996 linked first-change commits across 12,944 GitHub repositories. 
We found that generative AI is most frequently used during code implementation, placing its involvement at the point where new functionality is introduced into a project. 
However, this initial integration is frequently followed by refactoring, fixes, and structural adjustments, underscoring the continued role of developers in aligning generated output with project-specific constraints, quality standards, and evolving requirements. 
Over time, we also observed a gradual shift in how developers are engaging with AI. While code implementation remains dominant, developers are increasingly using AI for conceptual clarification, reasoning, and refining existing implementations. This trend suggests that AI-assisted development is an evolving, developer-driven workflow in which human oversight remains central, and value emerges through continued refinement rather than one-time code generation.

Our immediate future plan is to extend analysis beyond \textbf{first-change commits} to capture the long-term evolution of AI-assisted code, including stabilization, refactoring, and repeated modification patterns.
We further plan to expand the study beyond Python and JavaScript to examine whether the observed integration and adaptation behaviors generalize across additional programming languages, ecosystems, and development contexts. 
In addition, we plan to characterize the types of knowledge and conceptual support developers seek from AI and assess how effectively these needs are currently addressed. Lastly, we will study the sustained post-integration modifications to identify the gap between developer needs and the support provided by AI.

\section{Declarations}

\subsection{Funding: No funding was received to assist with the preparation of this manuscript.}
\subsection{Ethical Approval: Not Applicable. All publicly available data.}
\subsection{Informed consent: Not Applicable.}
\subsection{Author Contributions}
Abdullah Al Mujahid, Preetha Chatterjee, and Mia Mohammad Imran contributed to the conceptualization of the study. Abdullah Al Mujahid designed the methodology and conducted the primary analysis. Abdullah Al Mujahid and Mia Mohammad Imran contributed to the formal analysis and investigation. Abdullah Al Mujahid prepared the original manuscript draft. Abdullah Al Mujahid, Preetha Chatterjee, and Mia Mohammad Imran contributed to reviewing and editing the manuscript. Mia Mohammad Imran supervised the work.
\subsection{Data Availability Statement}
The data analyzed in this study are publicly available from the sources described in the manuscript.
\subsection{Conflict of Interest}
The authors have no competing interests with analyzing, studying, or publishing this research.

\bibliographystyle{ACM-Reference-Format}  
\bibliography{references}

@inproceedings{gao2024bayesian,
  title     = {Bayesian Calibration of Win Rate Estimation with {LLM} Evaluators},
  author    = {Gao, Yicheng and Xu, Gonghan and Wang, Zhe and Cohan, Arman},
  booktitle = {Proceedings of the 2024 Conference on Empirical Methods in Natural Language Processing},
  address   = {Miami, Florida, USA},
  publisher = {Association for Computational Linguistics},
  year      = {2024},
  month     = nov,
  pages     = {4757--4769},
  doi       = {10.18653/v1/2024.emnlp-main.273},
  url       = {https://aclanthology.org/2024.emnlp-main.273/}
}

@article{Ehsani_EMSE,
author = {Ehsani, Ramtin and Pathak, Sakshi and Parra, Esteban and Haiduc, Sonia and Chatterjee, Preetha},
title = {What characteristics make ChatGPT effective for software issue resolution? An empirical study of task, project, and conversational signals in GitHub issues},
year = {2025},
issue_date = {Nov 2025},
publisher = {Kluwer Academic Publishers},
address = {USA},
volume = {31},
number = {1},
issn = {1382-3256},
url = {https://doi.org/10.1007/s10664-025-10745-8},
doi = {10.1007/s10664-025-10745-8},
journal = {Empirical Softw. Engg.},
month = nov,
numpages = {36}
}

@inproceedings{hindle2009automated,
  title        = {Automatic classification of large changes into maintenance categories},
  author       = {Hindle, Abram and German, Daniel M and Holt, Ric C and Godfrey, Michael W},
  booktitle    = {Proceedings of the 2009 IEEE International Conference on Program Comprehension (ICPC)},
  pages        = {30--39},
  year         = {2009},
  organization = {IEEE}
}

@article{dawid1979maximum,
  title   = {Maximum likelihood estimation of observer error-rates using the EM algorithm},
  author  = {Dawid, A. Philip and Skene, Allan M.},
  journal = {Journal of the Royal Statistical Society: Series C (Applied Statistics)},
  volume  = {28},
  number  = {1},
  pages   = {20--28},
  year    = {1979},
  publisher = {Wiley}
}

@inproceedings{whitehill2009whose,
  title     = {Whose vote should count more: Optimal integration of labels from labelers of unknown expertise},
  author    = {Whitehill, Jacob and Ruvolo, Paul and Wu, Tingfan and Bergsma, Jacob and Movellan, Javier R.},
  booktitle = {Advances in Neural Information Processing Systems (NeurIPS 22)},
  pages     = {2035--2043},
  year      = {2009}
}

@inproceedings{snow2008cheap,
  title     = {Cheap and fast---but is it good? Evaluating non-expert annotations for natural language tasks},
  author    = {Snow, Rion and O'Connor, Brendan and Jurafsky, Daniel and Ng, Andrew Y.},
  booktitle = {Proceedings of the 2008 Conference on Empirical Methods in Natural Language Processing (EMNLP)},
  pages     = {254--263},
  year      = {2008},
  address   = {Honolulu, HI},
  publisher = {Association for Computational Linguistics}
}

@inproceedings{hovy2013mace,
  title     = {Learning whom to trust with {MACE}},
  author    = {Hovy, Dirk and Berg-Kirkpatrick, Taylor and Vaswani, Ashish and Hovy, Eduard},
  booktitle = {Proceedings of NAACL-HLT},
  pages     = {1120--1130},
  year      = {2013}
}

@inproceedings{he2024if,
  title={If in a crowdsourced data annotation pipeline, a gpt-4},
  author={He, Zeyu and Huang, Chieh-Yang and Ding, Chien-Kuang Cornelia and Rohatgi, Shaurya and Huang, Ting-Hao Kenneth},
  booktitle={Proceedings of the 2024 CHI Conference on Human Factors in Computing Systems},
  pages={1--25},
  year={2024}
}

@inproceedings{yao2024bayesian,
  title={A bayesian approach towards crowdsourcing the truths from llms},
  author={Yao, Peiran and Mathew, Jerin George and Singh, Shehraj and Firmani, Donatella and Barbosa, Denilson},
  booktitle={NeurIPS 2024 Workshop on Bayesian Decision-making and Uncertainty},
  year={2024}
}

@article{ibrahim2025learning,
  title={Learning from crowdsourced noisy labels: A signal processing perspective},
  author={Ibrahim, Shahana and Traganitis, Panagiotis A and Fu, Xiao and Giannakis, Georgios B},
  journal={IEEE Signal Processing Magazine},
  volume={42},
  number={3},
  pages={84--106},
  year={2025},
  publisher={IEEE}
}

@book{gelman2013bda3,
  title     = {Bayesian Data Analysis},
  author    = {Gelman, Andrew and Carlin, John B. and Stern, Hal S. and Dunson, David B. and Vehtari, Aki and Rubin, Donald B.},
  edition   = {3rd},
  publisher = {CRC Press},
  year      = {2013}
}

@article{rani2023decade,
  title={A decade of code comment quality assessment: A systematic literature review},
  author={Rani, Pooja and Blasi, Arianna and Stulova, Nataliia and Panichella, Sebastiano and Gorla, Alessandra and Nierstrasz, Oscar},
  journal={Journal of Systems and Software},
  volume={195},
  pages={111515},
  year={2023},
  publisher={Elsevier}
}

@article{quinn2018understanding,
  title={Understanding sequencing data as compositions: an outlook and review},
  author={Quinn, Thomas P and Erb, Ionas and Richardson, Mark F and Crowley, Tamsyn M},
  journal={Bioinformatics},
  volume={34},
  number={16},
  pages={2870--2878},
  year={2018},
  publisher={Oxford University Press}
}

@article{chandola2009anomaly,
  title={Anomaly detection: A survey},
  author={Chandola, Varun and Banerjee, Arindam and Kumar, Vipin},
  journal={ACM computing surveys (CSUR)},
  volume={41},
  number={3},
  pages={1--58},
  year={2009},
  publisher={ACM New York, NY, USA}
}

@book{box2015time,
  title={Time series analysis: forecasting and control},
  author={Box, George EP and Jenkins, Gwilym M and Reinsel, Gregory C and Ljung, Greta M},
  year={2015},
  publisher={John Wiley \& Sons}
}

@book{tukey1977exploratory,
  title={Exploratory data analysis},
  author={Tukey, John Wilder and others},
  volume={2},
  year={1977},
  publisher={Springer}
}

@article{mcinnes2017hdbscan,
  title     = {hdbscan: Hierarchical density based clustering},
  author    = {McInnes, Leland and Healy, John and Astels, Steve},
  journal   = {Journal of Open Source Software},
  volume    = {2},
  number    = {11},
  pages     = {205},
  year      = {2017},
  publisher = {The Open Journal},
  doi       = {10.21105/joss.00205},
  url       = {https://doi.org/10.21105/joss.00205}
}

@misc{octoverse2024blog,
  title  = {Octoverse 2024: AI leads Python to top language},
  author = {{GitHub}},
  year   = {2024},
  month  = {October},
  url    = {https://github.blog/news-insights/octoverse/octoverse-2024/},
  note   = {Available Online}
}

@misc{octoverse2023,
  title  = {The State of Open Source and AI in 2023},
  author = {{GitHub}},
  year   = {2023},
  month  = {November},
  url    = {https://github.blog/news-insights/research/the-state-of-open-source-and-ai/},
  note   = {Available Online}
}

@misc{so2024tech,
  title  = {Technology | 2024 Stack Overflow Developer Survey},
  author = {{Stack Overflow}},
  year   = {2024},
  url    = {https://survey.stackoverflow.co/2024/technology},
  note   = {Available Online}
}

@article{hao2024sharedconversations,
  title={An Empirical Study on Developers’ Shared Conversations with ChatGPT in GitHub Pull Requests and Issues},
  author={Hao, Huizi and Hasan, Kazi Amit and Qin, Hong and Macedo, Marcos and Tian, Yuan and Ding, Steven H. H. and Hassan, Ahmed E.},
  journal={Empirical Software Engineering},
  year={2024}
}

@inproceedings{mohamed2024chatting,
  title={Chatting with AI: Deciphering Developer Conversations with ChatGPT},
  author={Mohamed, Suad and Parvin, Abdullah and Parra, Esteban},
  booktitle={Proceedings of the 21st International Conference on Mining Software Repositories (MSR)},
  year={2024},
  publisher={ACM}
}

@inproceedings{sagdic2024taxonomy,
  title={On the Taxonomy of Developers’ Discussion Topics with ChatGPT},
  author={Sagdic, Ertugrul and Bayram, Arda and Islam, Md Rakibul},
  booktitle={Proceedings of the 21st International Conference on Mining Software Repositories (MSR) – Mining Challenge Track},
  year={2024},
  publisher={ACM}
}

@inproceedings{jin2024canchatgpt,
  title={Can chatgpt support developers? an empirical evaluation of large language models for code generation},
  author={Jin, Kailun and Wang, Chung-Yu and Pham, Hung Viet and Hemmati, Hadi},
  booktitle={Proceedings of the 21st International Conference on Mining Software Repositories},
  pages={167--171},
  year={2024}
}

@incollection{codabux2024teaching,
  title={Teaching Mining Software Repositories},
  author={Codabux, Zadia and Fard, Fatemeh and Verdecchia, Roberto and Palomba, Fabio and Di Nucci, Dario and Recupito, Gilberto},
  booktitle={Handbook on Teaching Empirical Software Engineering},
  pages={325--362},
  year={2024},
  publisher={Springer}
}

@inproceedings{mu2023developer,
  title={Developer-intent driven code comment generation},
  author={Mu, Fangwen and Chen, Xiao and Shi, Lin and Wang, Song and Wang, Qing},
  booktitle={2023 IEEE/ACM 45th International Conference on Software Engineering (ICSE)},
  pages={768--780},
  year={2023},
  organization={IEEE}
}

@article{kagdi2007survey,
  title={A survey and taxonomy of approaches for mining software repositories in the context of software evolution},
  author={Kagdi, Huzefa and Collard, Michael L and Maletic, Jonathan I},
  journal={Journal of software maintenance and evolution: Research and practice},
  volume={19},
  number={2},
  pages={77--131},
  year={2007},
  publisher={Wiley Online Library}
}

@inproceedings{steidl2013quality,
  title={Quality analysis of source code comments},
  author={Steidl, Daniela and Hummel, Benjamin and Juergens, Elmar},
  booktitle={2013 21st international conference on program comprehension (icpc)},
  pages={83--92},
  year={2013},
  organization={Ieee}
}

@inproceedings{hindle2008large,
  title={What do large commits tell us? a taxonomical study of large commits},
  author={Hindle, Abram and German, Daniel M and Holt, Ric},
  booktitle={Proceedings of the 2008 international working conference on Mining software repositories},
  pages={99--108},
  year={2008}
}

@article{barke2023grounded,
  title={Grounded copilot: How programmers interact with code-generating models},
  author={Barke, Shraddha and James, Michael B and Polikarpova, Nadia},
  journal={Proceedings of the ACM on Programming Languages},
  volume={7},
  number={OOPSLA1},
  pages={85--111},
  year={2023},
  publisher={ACM New York, NY, USA}
}

@inproceedings{alomar2024refactor,
  title={How to refactor this code? an exploratory study on developer-chatgpt refactoring conversations},
  author={AlOmar, Eman Abdullah and Venkatakrishnan, Anushkrishna and Mkaouer, Mohamed Wiem and Newman, Christian and Ouni, Ali},
  booktitle={Proceedings of the 21st International Conference on Mining Software Repositories},
  pages={202--206},
  year={2024}
}

@inproceedings{dvivedi2024comparative,
  title={A comparative analysis of large language models for code documentation generation},
  author={Dvivedi, Shubhang Shekhar and Vijay, Vyshnav and Pujari, Sai Leela Rahul and Lodh, Shoumik and Kumar, Dhruv},
  booktitle={Proceedings of the 1st ACM international conference on AI-powered software},
  pages={65--73},
  year={2024}
}

@article{murali2024ai,
  title={AI-assisted Code Authoring at Scale: Fine-tuning, deploying, and mixed methods evaluation},
  author={Murali, Vijayaraghavan and Maddila, Chandra and Ahmad, Imad and Bolin, Michael and Cheng, Daniel and Ghorbani, Negar and Fernandez, Renuka and Nagappan, Nachiappan and Rigby, Peter C},
  journal={Proceedings of the ACM on Software Engineering},
  volume={1},
  number={FSE},
  pages={1066--1085},
  year={2024},
  publisher={ACM New York, NY, USA}
}

@inproceedings{mozannar2024reading,
  title={Reading between the lines: Modeling user behavior and costs in AI-assisted programming},
  author={Mozannar, Hussein and Bansal, Gagan and Fourney, Adam and Horvitz, Eric},
  booktitle={Proceedings of the 2024 CHI Conference on Human Factors in Computing Systems},
  pages={1--16},
  year={2024}
}

@inproceedings{du2024evaluating,
  title={Evaluating large language models in class-level code generation},
  author={Du, Xueying and Liu, Mingwei and Wang, Kaixin and Wang, Hanlin and Liu, Junwei and Chen, Yixuan and Feng, Jiayi and Sha, Chaofeng and Peng, Xin and Lou, Yiling},
  booktitle={Proceedings of the IEEE/ACM 46th International Conference on Software Engineering},
  pages={1--13},
  year={2024}
}

@article{peng2023impact,
  title={The impact of ai on developer productivity: Evidence from github copilot},
  author={Peng, Sida and Kalliamvakou, Eirini and Cihon, Peter and Demirer, Mert},
  journal={arXiv preprint arXiv:2302.06590},
  year={2023}
}

@inproceedings{chouchen2024software,
  title={How do software developers use chatgpt? an exploratory study on github pull requests},
  author={Chouchen, Moataz and Bessghaier, Narjes and Begoug, Mahi and Ouni, Ali and Alomar, Eman and Mkaouer, Mohamed Wiem},
  booktitle={Proceedings of the 21st International Conference on Mining Software Repositories},
  pages={212--216},
  year={2024}
}

@inproceedings{grewal2024analyzing,
  title={Analyzing developer use of chatgpt generated code in open source github projects},
  author={Grewal, Balreet and Lu, Wentao and Nadi, Sarah and Bezemer, Cor-Paul},
  booktitle={Proceedings of the 21st International Conference on Mining Software Repositories},
  pages={157--161},
  year={2024}
}

@inproceedings{guo2024exploring,
  title={Exploring the potential of chatgpt in automated code refinement: An empirical study},
  author={Guo, Qi and Cao, Junming and Xie, Xiaofei and Liu, Shangqing and Li, Xiaohong and Chen, Bihuan and Peng, Xin},
  booktitle={Proceedings of the 46th IEEE/ACM International Conference on Software Engineering},
  pages={1--13},
  year={2024}
}

@software{replication-package,
  title        = {Replication Package},
  author       = {Anonymous},
  year         = {2026},
  url          = {https://github.com/MSwadhin/empirical-study-dev-ai-usage},
  howpublished = {\url{https://github.com/MSwadhin/empirical-study-dev-ai-usage}}
}

@inproceedings{hu2022practitioners,
  title={Practitioners' expectations on automated code comment generation},
  author={Hu, Xing and Xia, Xin and Lo, David and Wan, Zhiyuan and Chen, Qiuyuan and Zimmermann, Thomas},
  booktitle={Proceedings of the 44th international conference on software engineering},
  pages={1693--1705},
  year={2022}
}

@article{xue2024automated,
  title={Automated commit message generation with large language models: An empirical study and beyond},
  author={Xue, Pengyu and Wu, Linhao and Yu, Zhongxing and Jin, Zhi and Yang, Zhen and Li, Xinyi and Yang, Zhenyu and Tan, Yue},
  journal={IEEE Transactions on Software Engineering},
  year={2024},
  publisher={IEEE}
}

@inproceedings{ehsani2025towards,
  title={Towards detecting prompt knowledge gaps for improved llm-guided issue resolution},
  author={Ehsani, Ramtin and Pathak, Sakshi and Chatterjee, Preetha},
  booktitle={2025 IEEE/ACM 22nd International Conference on Mining Software Repositories (MSR)},
  pages={699--711},
  year={2025},
  organization={IEEE}
}

@inproceedings{katzy2025Qualitative,
author = {Katzy, Jonathan and Huang, Yongcheng and Panchu, Gopal-Raj and Ziemlewski, Maksym and Loizides, Paris and Vermeulen, Sander and van Deursen, Arie and Izadi, Maliheh},
title = {A Qualitative Investigation into LLM-Generated Multilingual Code Comments and Automatic Evaluation Metrics},
year = {2025},
isbn = {9798400715945},
publisher = {Association for Computing Machinery},
address = {New York, NY, USA},
booktitle = {Proceedings of the 21st International Conference on Predictive Models and Data Analytics in Software Engineering},
pages = {31–40},
numpages = {10},
location = {Trondheim, Norway},
series = {PROMISE '25}
}

@article{li2026comments,
title = {Do comments and expertise still matter? An experiment on programmers’ adoption of AI-generated JavaScript code},
journal = {Journal of Systems and Software},
volume = {231},
pages = {112634},
year = {2026},
issn = {0164-1212},
author = {Changwen Li and Christoph Treude and Ofir Turel},
}

@incollection{tan2015code,
  title={Code comment analysis for improving software quality},
  author={Tan, Lin},
  booktitle={The art and science of analyzing software data},
  pages={493--517},
  year={2015},
  publisher={Elsevier}
}

@misc{ifeedback_sample_size_calculator,
  author       = {{iFeedback}},
  title        = {Sample Size Calculator},
  year         = {2026},
  url          = {https://ifeedback.co.za/resources/sample-size-calculator},
  note         = {Available Online}
}

@misc{geopoll_sample_size_research,
  author       = {{GeoPoll}},
  title        = {What Is the Right Sample Size for Research?},
  year         = {2021},
  url          = {https://www.geopoll.com/blog/sample-size-research},
  note         = {Available Online}
}

@inproceedings{jimenez2023swe,
  title={Swe-bench: Can language models resolve real-world github issues?},
  author={Jimenez, Carlos E and Yang, John and Wettig, Alexander and Yao, Shunyu and Pei, Kexin and Press, Ofir and Narasimhan, Karthik},
  booktitle={International Conference on Learning Representations},
  volume={2024},
  pages={54107--54157},
  year={2024}
}

@inproceedings{ferreira2022characterizing,
  title={Characterizing commits in open-source software},
  author={Ferreira, M{\'\i}vian and Gon{\c{c}}alves, Diego and Bigonha, Mariza and Ferreira, Kecia},
  booktitle={Proceedings of the XXI Brazilian Symposium on Software Quality},
  pages={1--10},
  year={2022}
}

@inproceedings{tian2022makes,
  title={What makes a good commit message?},
  author={Tian, Yingchen and Zhang, Yuxia and Stol, Klaas-Jan and Jiang, Lin and Liu, Hui},
  booktitle={Proceedings of the 44th International Conference on Software Engineering},
  pages={2389--2401},
  year={2022}
}

@misc{grootendorst_bertopic_algorithm,
  author       = {Grootendorst, Maarten},
  title        = {BERTopic: Algorithm},
  year         = {n.d.},
  howpublished = {\url{https://maartengr.github.io/BERTopic/algorithm/algorithm.html}},
  note         = {Accessed: 19 December 2025}
}

@article{grootendorst2022bertopic,
  title={BERTopic: Neural topic modeling with a class-based TF-IDF procedure},
  author={Grootendorst, Maarten},
  journal={arXiv preprint arXiv:2203.05794},
  year={2022}
}

@inproceedings{yamauchi2014clustering,
  title={Clustering commits for understanding the intents of implementation},
  author={Yamauchi, Kenji and Yang, Jiachen and Hotta, Keisuke and Higo, Yoshiki and Kusumoto, Shinji},
  booktitle={2014 IEEE international conference on software maintenance and evolution},
  pages={406--410},
  year={2014},
  organization={IEEE}
}

@inproceedings{fan2023large,
  title={Large language models for software engineering: Survey and open problems},
  author={Fan, Angela and Gokkaya, Beliz and Harman, Mark and Lyubarskiy, Mitya and Sengupta, Shubho and Yoo, Shin and Zhang, Jie M},
  booktitle={2023 IEEE/ACM International Conference on Software Engineering: Future of Software Engineering (ICSE-FoSE)},
  pages={31--53},
  year={2023},
  organization={IEEE}
}

@inproceedings{aguiar2024multi,
  title={Multi-language software development in the llm era: Insights from practitioners’ conversations with chatgpt},
  author={Aguiar, Lucas and Paixao, Matheus and Carmo, Rafael and Soares, Edson and Leal, Antonio and Freitas, Matheus and Gama, Eliakim},
  booktitle={Proceedings of the 18th ACM/IEEE International Symposium on Empirical Software Engineering and Measurement},
  pages={489--495},
  year={2024}
}

@article{ozkaya2023application,
  title={Application of large language models to software engineering tasks: Opportunities, risks, and implications},
  author={Ozkaya, Ipek},
  journal={IEEE Software},
  volume={40},
  number={3},
  pages={4--8},
  year={2023},
  publisher={IEEE}
}

@inproceedings{belzner2023large,
  title={Large language model assisted software engineering: prospects, challenges, and a case study},
  author={Belzner, Lenz and Gabor, Thomas and Wirsing, Martin},
  booktitle={International conference on bridging the gap between AI and reality},
  pages={355--374},
  year={2023},
  organization={Springer}
}

@inproceedings{moroz2022potential,
  title={The potential of artificial intelligence as a method of software developer's productivity improvement},
  author={Moroz, Ekaterina A and Grizkevich, Vladimir O and Novozhilov, Igor M},
  booktitle={2022 Conference of Russian Young Researchers in Electrical and Electronic Engineering (ElConRus)},
  pages={386--390},
  year={2022},
  organization={IEEE}
}

@article{gao2025current,
  title={The current challenges of software engineering in the era of large language models},
  author={Gao, Cuiyun and Hu, Xing and Gao, Shan and Xia, Xin and Jin, Zhi},
  journal={ACM Transactions on Software Engineering and Methodology},
  volume={34},
  number={5},
  pages={1--30},
  year={2025},
  publisher={ACM New York, NY}
}

@misc{CDC_StatisticalSignificance_2024,
  author       = {{CDC}},
  title        = {Centers for Disease Control and Prevention, National Center for Health Statistics: Statistical significance},
  year         = {2024},
  howpublished = {\url{https://www.cdc.gov/nchs/hus/sources-definitions/statistical-significance.htm}},
  note         = {Last reviewed July 30, 2024},
}

@article{mcinnes2018umap,
  title={Umap: Uniform manifold approximation and projection for dimension reduction},
  author={McInnes, Leland and Healy, John and Melville, James},
  journal={arXiv preprint arXiv:1802.03426},
  year={2018}
}

@inproceedings{muennighoff2023mteb,
  title={Mteb: Massive text embedding benchmark},
  author={Muennighoff, Niklas and Tazi, Nouamane and Magne, Lo{\"\i}c and Reimers, Nils},
  booktitle={Proceedings of the 17th Conference of the European Chapter of the Association for Computational Linguistics},
  pages={2014--2037},
  year={2023}
}

@inproceedings{hassan2008automated,
  title={Automated classification of change messages in open source projects},
  author={Hassan, Ahmed E},
  booktitle={Proceedings of the 2008 ACM symposium on Applied computing},
  pages={837--841},
  year={2008}
}

@inproceedings{tian2012information,
  title={Information retrieval based nearest neighbor classification for fine-grained bug severity prediction},
  author={Tian, Yuan and Lo, David and Sun, Chengnian},
  booktitle={2012 19th Working Conference on Reverse Engineering},
  pages={215--224},
  year={2012},
  organization={IEEE}
}

@article{imran2026olaf,
  title={OLAF: Towards Robust LLM-Based Annotation Framework in Empirical Software Engineering},
  author={Imran, Mia Mohammad and Zaman, Tarannum Shaila},
  journal={3rd International Workshop on Methodological Issues with Empirical Studies in Software Engineering (WSESE)},
  year={2026}
}

@inproceedings{mujahid2026genai_satd,
  title={" TODO: Fix the Mess Gemini Created": Towards Understanding GenAI-Induced Self-Admitted Technical Debt},
  author={Mujahid, Abdullah Al and Imran, Mia Mohammad},
  booktitle = {Proceedings of the 9th International Conference on Technical Debt},
  year      = {2026},
}

@misc{ruppert2004elements,
  title={The elements of statistical learning: data mining, inference, and prediction},
  author={Ruppert, David},
  year={2004},
  publisher={Taylor \& Francis}
}

@inproceedings{abdelsalam2026humansLLMs,
  author    = {Youssef Abdelsalam and Norman Peitek and Anna-Maria Maurer and Mariya Toneva and Sven Apel},
  title     = {Are Humans and LLMs Confused by the Same Code? An Empirical Study on Fixation-Related Potentials and LLM Perplexity},
  booktitle = {2026 IEEE/ACM 45th International Conference on Software Engineering (ICSE)},
  year      = {2026},
  organization={IEEE}
}

@misc{li2025aidev,
  title        = {The Rise of AI Teammates in Software Engineering (SE) 3.0: How Autonomous Coding Agents Are Reshaping Software Engineering},
  author       = {Hao Li and Haoxiang Zhang and Ahmed E. Hassan},
  year         = {2025},
  eprint       = {2507.15003},
  archivePrefix= {arXiv},
  primaryClass = {cs.SE},
  url          = {https://arxiv.org/abs/2507.15003}
}

@article{saunders2018saturation,
  title={Saturation in qualitative research: exploring its conceptualization and operationalization},
  author={Saunders, Benjamin and Sim, Julius and Kingstone, Tom and Baker, Shula and Waterfield, Jackie and Bartlam, Bernadette and Burroughs, Heather and Jinks, Clare},
  journal={Quality \& quantity},
  volume={52},
  number={4},
  pages={1893--1907},
  year={2018},
  publisher={Springer}
}

@article{maguire2017doing,
  title={Doing a thematic analysis: A practical, step-by-step guide for learning and teaching scholars.},
  author={Maguire, Moira and Delahunt, Brid},
  journal={All Ireland journal of higher education},
  volume={9},
  number={3},
  year={2017}
}

@inproceedings{ackerman1998considering,
  title={Considering an organization's memory},
  author={Ackerman, Mark S and Halverson, Christine},
  booktitle={Proceedings of the 1998 ACM conference on Computer supported cooperative work},
  pages={39--48},
  year={1998}
}

@article{mohamed2025impact,
  title={The impact of LLM-assistants on software developer productivity: A systematic literature review},
  author={Mohamed, Amr and Assi, Maram and Guizani, Mariam},
  journal={arXiv preprint arXiv:2507.03156},
  year={2025}
}

@inproceedings{weisz2025examining,
  title={Examining the use and impact of an ai code assistant on developer productivity and experience in the enterprise},
  author={Weisz, Justin D and Kumar, Shraddha Vijay and Muller, Michael and Browne, Karen-Ellen and Goldberg, Arielle and Heintze, Katrin Ellice and Bajpai, Shagun},
  booktitle={Proceedings of the Extended Abstracts of the CHI Conference on Human Factors in Computing Systems},
  pages={1--13},
  year={2025}
}

@article{khojah2024beyond,
  title={Beyond code generation: An observational study of chatgpt usage in software engineering practice},
  author={Khojah, Ranim and Mohamad, Mazen and Leitner, Philipp and de Oliveira Neto, Francisco Gomes},
  journal={Proceedings of the ACM on Software Engineering},
  volume={1},
  number={FSE},
  pages={1819--1840},
  year={2024},
  publisher={ACM New York, NY, USA}
}

@inproceedings{contreras2024conversational,
  title={Conversational Assistants for Software Development: Integration, Traceability and Coordination.},
  author={Contreras, Albert and Guerra, Esther and de Lara, Juan},
  booktitle={ENASE},
  pages={27--38},
  year={2024}
}

@incollection{rogers2014diffusion,
  title={Diffusion of innovations},
  author={Rogers, Everett M and Singhal, Arvind and Quinlan, Margaret M},
  booktitle={An integrated approach to communication theory and research},
  pages={432--448},
  year={2014},
  publisher={Routledge}
}

@article{mieritz2005defining,
  title={Defining Gartner total cost of ownership},
  author={Mieritz, Lars and Kirwin, Bill},
  journal={L. Mieritz, B. Kirwin},
  year={2005}
}

@book{gwet2014handbook,
  title={Handbook of inter-rater reliability: The definitive guide to measuring the extent of agreement among raters},
  author={Gwet, Kilem L},
  year={2014},
  publisher={Advanced Analytics, LLC}
}

@article{walsh2022assessing,
  title={Assessing interrater reliability of a faculty-provided feedback rating instrument},
  author={Walsh, Daniel P and Chen, Michael J and Buhl, Lauren K and Neves, Sara E and Mitchell, John D},
  journal={Journal of Medical Education and Curricular Development},
  volume={9},
  pages={23821205221093205},
  year={2022},
  publisher={SAGE Publications Sage UK: London, England}
}

@article{wongpakaran2013comparison,
  title={A comparison of Cohen’s Kappa and Gwet’s AC1 when calculating inter-rater reliability coefficients: a study conducted with personality disorder samples},
  author={Wongpakaran, Nahathai and Wongpakaran, Tinakon and Wedding, Danny and Gwet, Kilem L},
  journal={BMC medical research methodology},
  volume={13},
  number={1},
  pages={61},
  year={2013},
  publisher={Springer}
}

@article{huang2026issuecommitllm,
  title={Back to the Basics: Rethinking Issue-Commit Linking with LLM-Assisted Retrieval},
  author={Huang, Huihui and Widyasari, Ratnadira and Zhang, Ting and Irsan, Ivana Clairine and Shi, Jieke and Ang, Han Wei and Liauw, Frank and Ouh, Eng Lieh and Shar, Lwin Khin and Kang, Hong Jin and others},
  journal={arXiv preprint arXiv:2507.09199},
  year={2025}
}

@inproceedings{khatib2026assertflip,
  title={AssertFlip: Reproducing Bugs via Inversion of LLM-Generated Passing Tests},
  author={Khatib, Lara and Mathews, Noble Saji and Nagappan, Mei},
  booktitle={2026 IEEE/ACM International Conference on Software Engineering(ICSE)},
  year={2026},
  organization={IEEE}
}

@inproceedings{gu2026semanticrepairllm,
  title={A Semantic-based Optimization Approach for Repairing LLMs: Case Study on Code Generation},
  author={Gu, Jian and Aleti, Aldeida and Chen, Chunyang and Zhang, Hongyu},
  booktitle={2026 IEEE/ACM International Conference on Software Engineering(ICSE)},
  year={2026},
  organization={IEEE}
}

@article{newey1986simple,
  title={A simple, positive semi-definite, heteroskedasticity and autocorrelationconsistent covariance matrix},
  author={Newey, Whitney K and West, Kenneth D},
  year={1986},
  publisher={National Bureau of Economic Research Cambridge, Mass., USA}
}

@article{chandiramani2026nemotron,
  title={Nemotron 3 Super: Open, Efficient Mixture-of-Experts Hybrid Mamba-Transformer Model for Agentic Reasoning},
  author={Chandiramani, Aakshita and Blakeman, Aaron and Olaoye, Abdullahi and Gupta, Abhibha and Somasamudramath, Abhilash and Khattar, Abhinav and Adesoba, Adeola and Renduchintala, Adi and Asif, Adil and Agrawal, Aditya and others},
  journal={arXiv preprint arXiv:2604.12374},
  year={2026}
}

@misc{gemma4_website,
  author       = {{Google DeepMind}},
  title        = {Gemma 4: Lightweight, state-of-the-art open models},
  year         = {2026},
  howpublished = {\url{https://blog.google/technology/developers/gemma-4/}},
  note         = {Accessed: 2026-04-30}
}

\end{document}